\documentclass[aps,prl,twocolumn,superscriptaddress]{revtex4}
\usepackage{mathrsfs}
\usepackage{epsfig}
\usepackage{graphicx}
\usepackage{amsfonts}
\usepackage[figuresright]{rotating}
\usepackage{amssymb}
\usepackage{amsmath}
\usepackage{dcolumn}
\usepackage{bm}
\usepackage{color}

\def\be{\begin{equation}} \def\ee{\end{equation}}
\def\bea{\begin{eqnarray}} \def\eea{\end{eqnarray}}

\newcommand{\WQC} {Wilczek Quantum Center and Key Laboratory of Artificial Structures and Quantum Control, School of Physics and Astronomy, Shanghai Jiao Tong University, Shanghai 200240, China}

\newcommand{\SRCQC}{Shanghai Research Center for Quantum Sciences, Shanghai 201315, China}

\begin{document}
\title{Localization of matter waves in lattice systems with moving disorder}

\author{Chenyue Guo }
\affiliation{\WQC}


\author{Zi Cai}
\email{zcai@sjtu.edu.cn}
\affiliation{\WQC}
\affiliation{\SRCQC}

\begin{abstract}  
We study the localization phenomena in a one-dimensional lattice system with a uniformly moving disordered potential. At a low moving velocity, we find a sliding localized phase in which the initially localized matter wave adiabatically follows the moving potential without diffusion, thus resulting in an initial state memory in the many-body dynamics. Such an intriguing localized phase distinguishes itself from the standard Anderson localization in two aspects: it is not robust against interaction, but persists in  the presence of slowly varying perturbations. Such a sliding localized phase can be understood as a consequence of interference between the wavepacket paths under moving quasi-periodic potentials with various periods that are incommensurate with the lattice constant. The experimental realization and detection were also discussed.

\end{abstract}


\maketitle

{\it Introduction --} The absence of the diffusion of waves in a disordered medium (dubbed Anderson localization)  originates from the interference
between various scattering paths, thus it is ubiquitous in wave physics\cite{Anderson1958}.  Such a phenomenon is expected to be robust against weak interactions, and has reattracted enormous interest recently in the context of many-body localization (MBL)\cite{Basko2006,Oganesyan2007,Znidaric2008,Pal2010}. Even though disorders naturally exist in solid-state settings, they can also be artificially introduced into intrinsically clean systems (e.g.,ultracold atom or trapped ion) in a controllable matter\cite{Clement2006,Bermudez2010}. Owing to its  unique features such as the perfect isolation and high degree of parameter tunability\cite{Bloch2008}, the cold-atom system represents a new perspective for studying  localization\cite{Fallani2007,Billy2008,Roati2008,Schreiber2015}. For instance, it allows the exploration of  localization physics in a far-from-equilibrium system within a strong driving regime  inaccessible in conventional solid-state settings\cite{Bordia2017,Vershinina2017,Liu2018}, and thus is beyond the scope of the  linear response theory derived by Mott\cite{Mott1968}.

In cold atom experiments, a quantum many-body system can be driven out of equilibrium by periodically or stochastically modulating the system parameters. Recently, an intriguing driving protocol other than a regular (periodic) or completely irregular (stochastic) driving protocols  was proposed\cite{Xu2018}. Therein, neither a spatial translational symmetry, nor a temporal translational symmetry (TTS) are present for the driving potential, which, instead, exhibits nontrivial intertwined space-time symmetries  that cannot be decomposed into a direct product of spatial and temporal symmetries. Owing to the absence of discrete TTS, such a driving protocol significantly differs from the periodic driving, thus the widely employed Floquet description no longer apply, nor it resembles the stochastic or quasi-periodic ones\cite{Dumitrescu2018,Cai2020,Long2022,Zhao2022} due to its strong correlation along a particular space-time direction.

\begin{figure}[htb]
\includegraphics[width=0.99\linewidth]{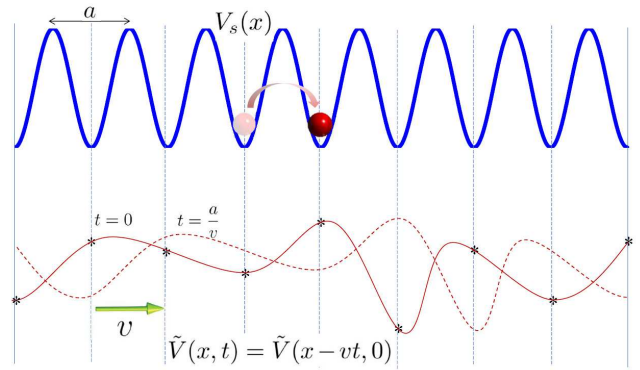}
\caption{(Color online) Schematic of the static periodic and moving disordered potentials in our model.}\label{fig:pic}
\end{figure}

Such an intriguing driving protocol avails  new possibilities exploring non-equilibrium physics beyond the scope of periodically driven systems\cite{Eckardt2017,Harper2020,Oka2019}. As a example, in this  study, we explore the localization physics in a driven-disordered system with a sliding space-time translational symmetry($\mathbf{SSTS}$). The proposed system is a one-dimensional (1D) quantum model subjected to a time-dependent inhomogeneous potential as a superposition of a dynamic disordered potential moving at a constant velocity and  a static periodic one (see Fig.\ref{fig:pic}). As would be demonstrated, such a moving disordered potential with tunable velocity can be readily achieved in the cold atomic systems, even though it is unrealistic in solid-state settings.  Contrary to the periodically-driven disordered systems which are generally delocalized by low-frequency perturbations\cite{Martinez2006,Ponte2015,Lazarides2015,Abanin2016,Vershinina2017}, in the proposed model, a sliding localized phase ($\mathbf{SLP}$) was observed  at a sufficiently low velocity  of the moving potential. Even though the center of mass (COM) of the matter wave adiabatically follows the moving potential, it does not cause diffusion.  Such an SLP persists in the non-interacting many-body systems, which exhibit an initial state ``memory'' effect. However, dissimilar  to the MBL systems, SLP will be delocalized even by weak interactions. The system is delocalized at a high moving velocity. Moreover, instead of diffusion, we find a ballistic transport  in the single-particle dynamics, and an algebraic decay of the initial state memory at the many-body level.

{\it Model --} First, we considered a time-dependent single-particle Hamiltonian, which is defined in the 1D continuum space as follows:
\begin{equation}
\hat{H}(x,t)=\frac{\hat{P}^2}{2m}+V_s(x)+ \tilde{V}(x,t)\label{eq:Ham1}
\end{equation}
where $V_s(x)=V_s(x+a)$ is a static periodic potential that is produced by the optical lattice with lattice constant $a$. $\tilde{V}(x,t)$ represents the time-dependent disordered potential that breaks the spatial translational symmetry at any given time. To drive the system out of equilibrium, the disordered potential is suddenly pushed forward at a constant velocity $\mathbf{v}$ indicating $\tilde{V}(x,t)=\tilde{V}(x-vt,0)$.   The explicit form of $\tilde{V}(x,t)$ would be formulated subsequently. $\hat{H}(x,t)$ preserves the SSTS even though it breaks the temporal and spacial translational symmetry separately:
\begin{equation}
\hat{H}(x,t)=\hat{H}(x+a,t+a/v). \label{eq:symmetry}
\end{equation}
 Provided the static periodic potential is sufficiently deep, we adopted the single-band approximation and derived the tight-binding Hamiltonian in the 1D lattice as:
\begin{equation}
\hat{H}(t)=\sum_i [-J(C_i^\dag C_{i+1}+h.c)+V_i(t) \hat{n}_i] \label{eq:Ham2}
\end{equation}
where $C_i$ ($C_i^\dag$) is the annihilation (creation) operator of the spinless fermion on site $i$ and $\hat{n}_i=C_i^\dag C_i$.  $J$ is the nearest-neighboring (NN) single-particle hopping amplitude and $V_i(t)=\tilde{V}(x=ai,t)$ is the moving disordered potential at time t. Assuming  $V_i(t=0)$ was randomly sampled from a uniform random distribution with $V_i(t=0)\in[-\Delta,\Delta]$.  To recover the continuum function $\tilde{V}(x,t=0)$ from a set of given points $\{V_i(t=0)\}$,  we performed cubic spline interpolation to smoothly connect these points on the different sites, and derived the continuum function $\tilde{V}(x,t=0)$.  The results did not significantly depend on specific interpolation methods \cite{Supplementary}. Immediately after determining $\tilde{V}(x,t=0)$, $V_i(t)$ at any given time can be obtained using the identities $\tilde{V}(x,t)=\tilde{V}(x-vt,0)$ and $V_i(t)=\tilde{V}(x=ai,t)$. Despite the similarity in their SSTS, we should emphasize that  the proposed potential is significantly different from that in the space-time crystal\cite{Xu2018,Gao2021}, since our potential has only one lattice vector as shown in Eq.(\ref{eq:symmetry}), thus it is not a crystal in the 1+1 D space-time.

\begin{figure*}[htb]
\includegraphics[width=0.325\linewidth,bb=58 58 609 537]{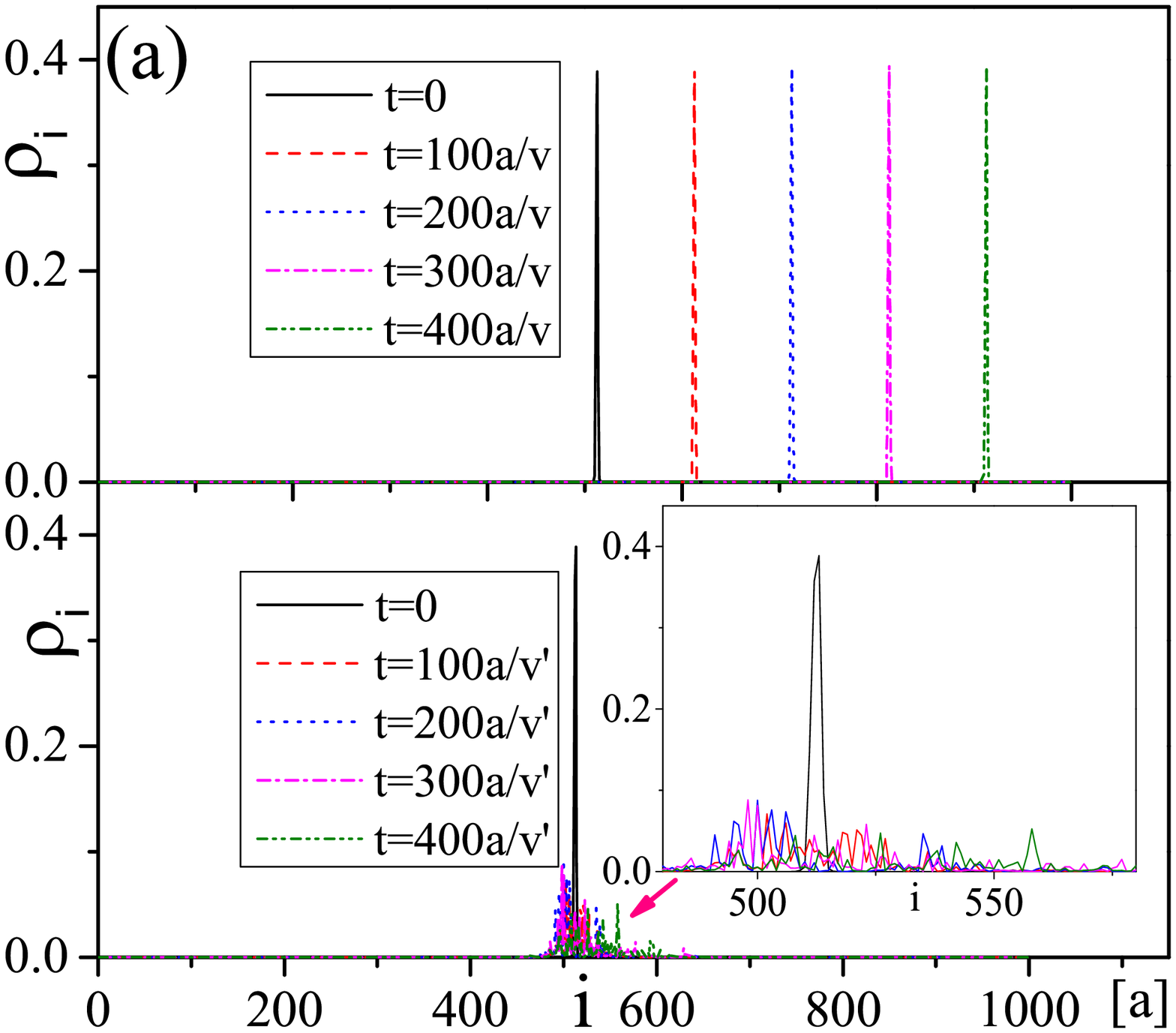}
\includegraphics[width=0.33\linewidth,bb=58 58 609 537]{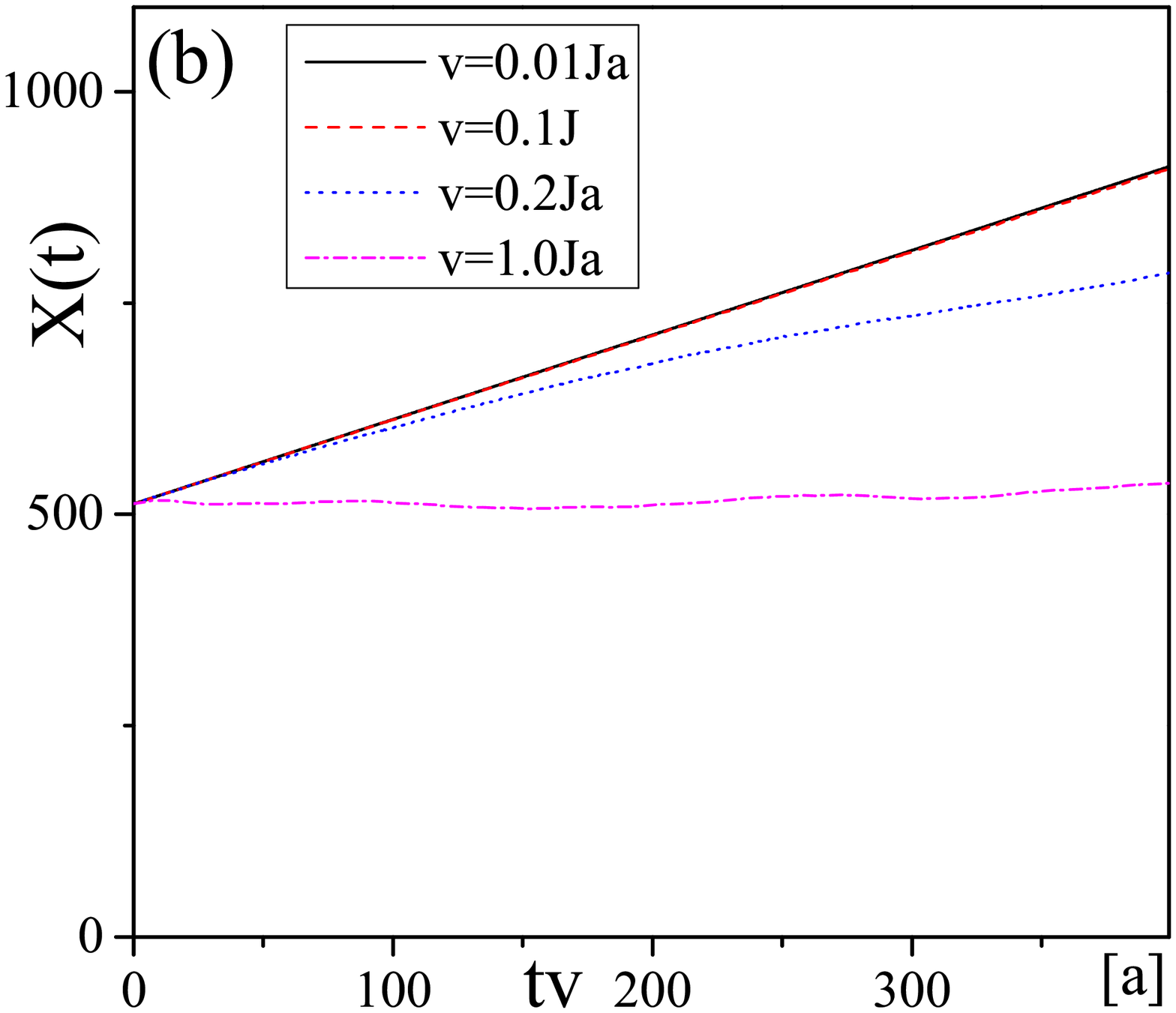}
\includegraphics[width=0.33\linewidth,bb=58 58 609 537]{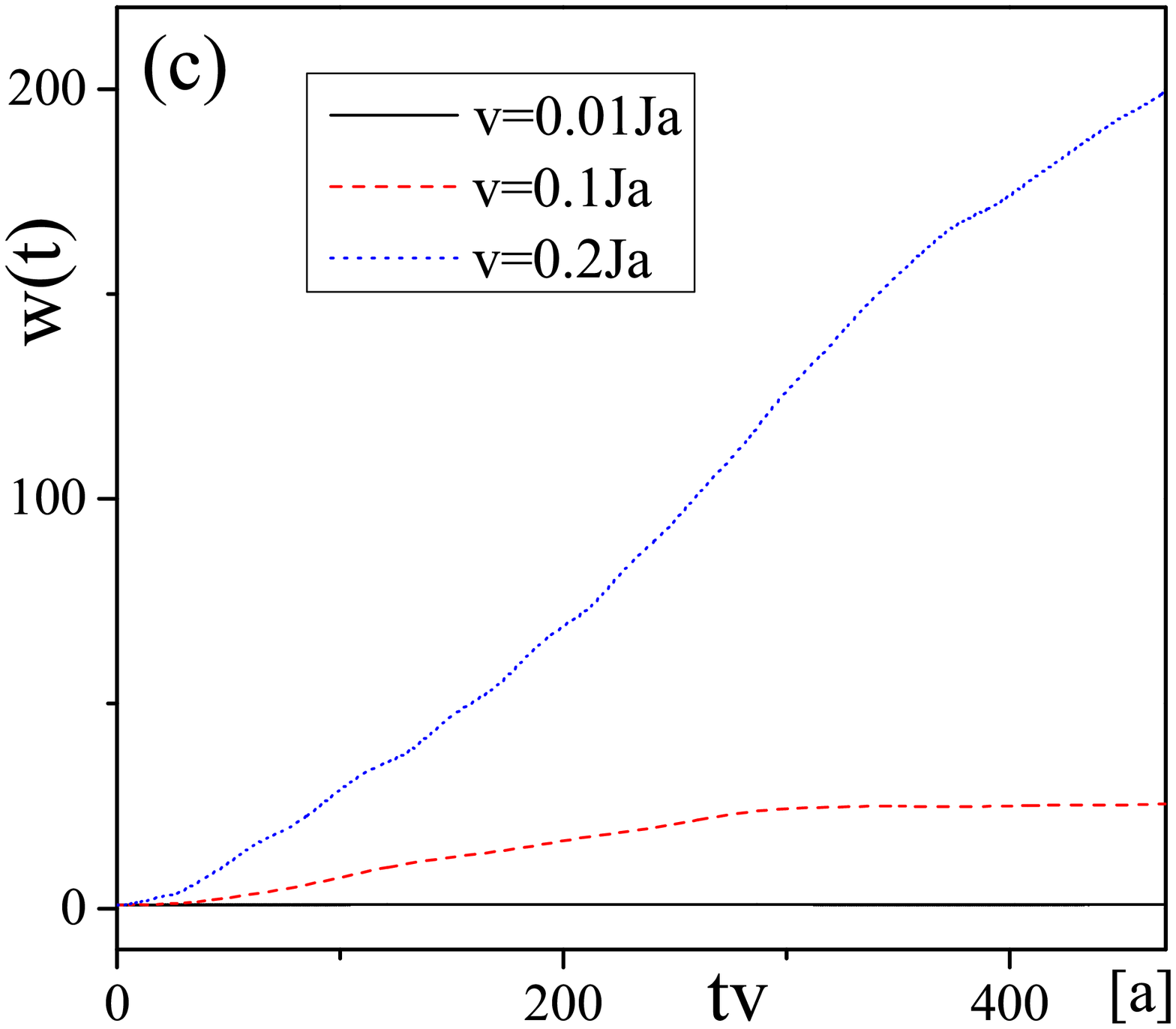}
\includegraphics[width=0.33\linewidth,bb=88 63 629 563]{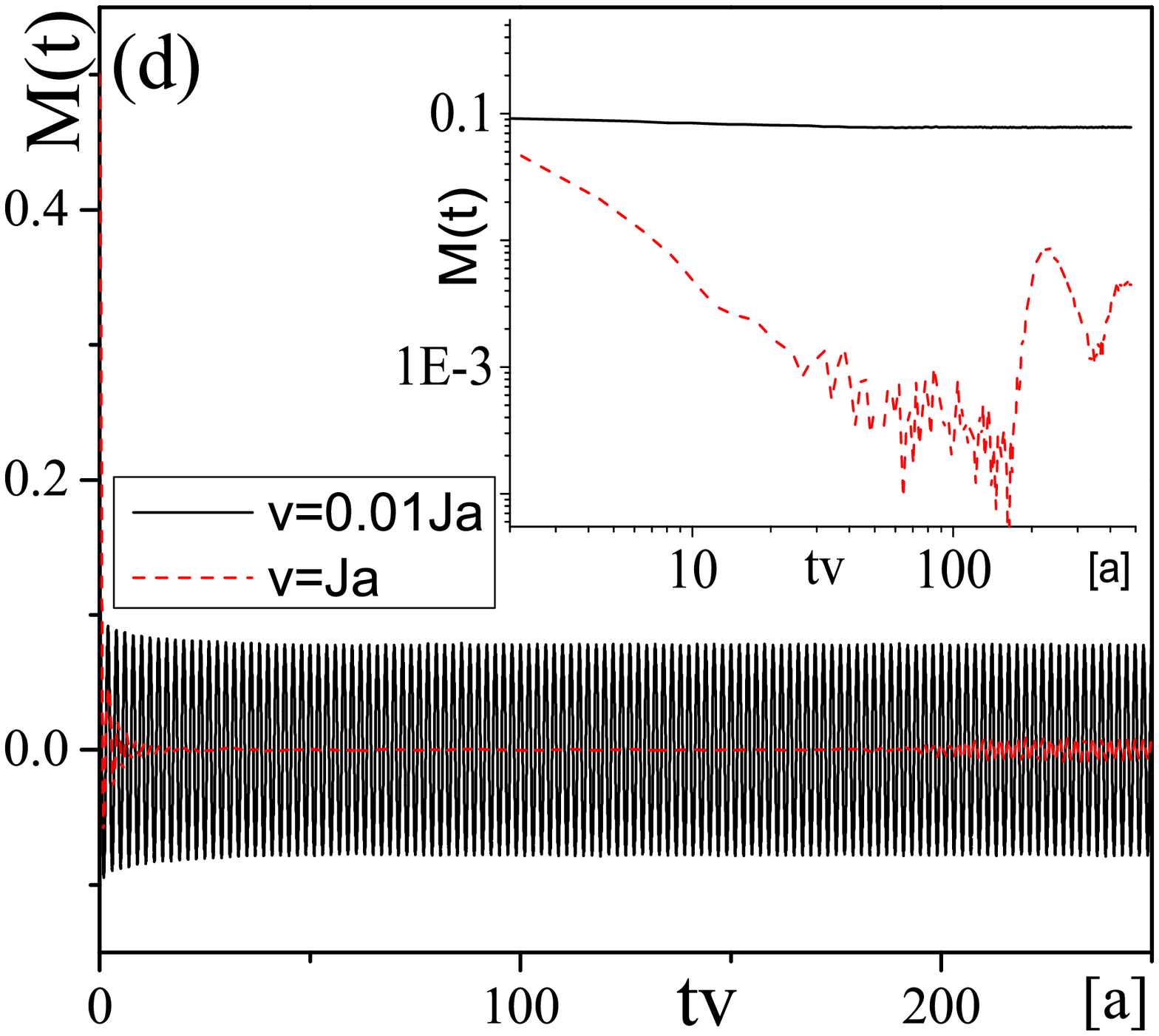}
\includegraphics[width=0.325\linewidth,bb=58 58 609 537]{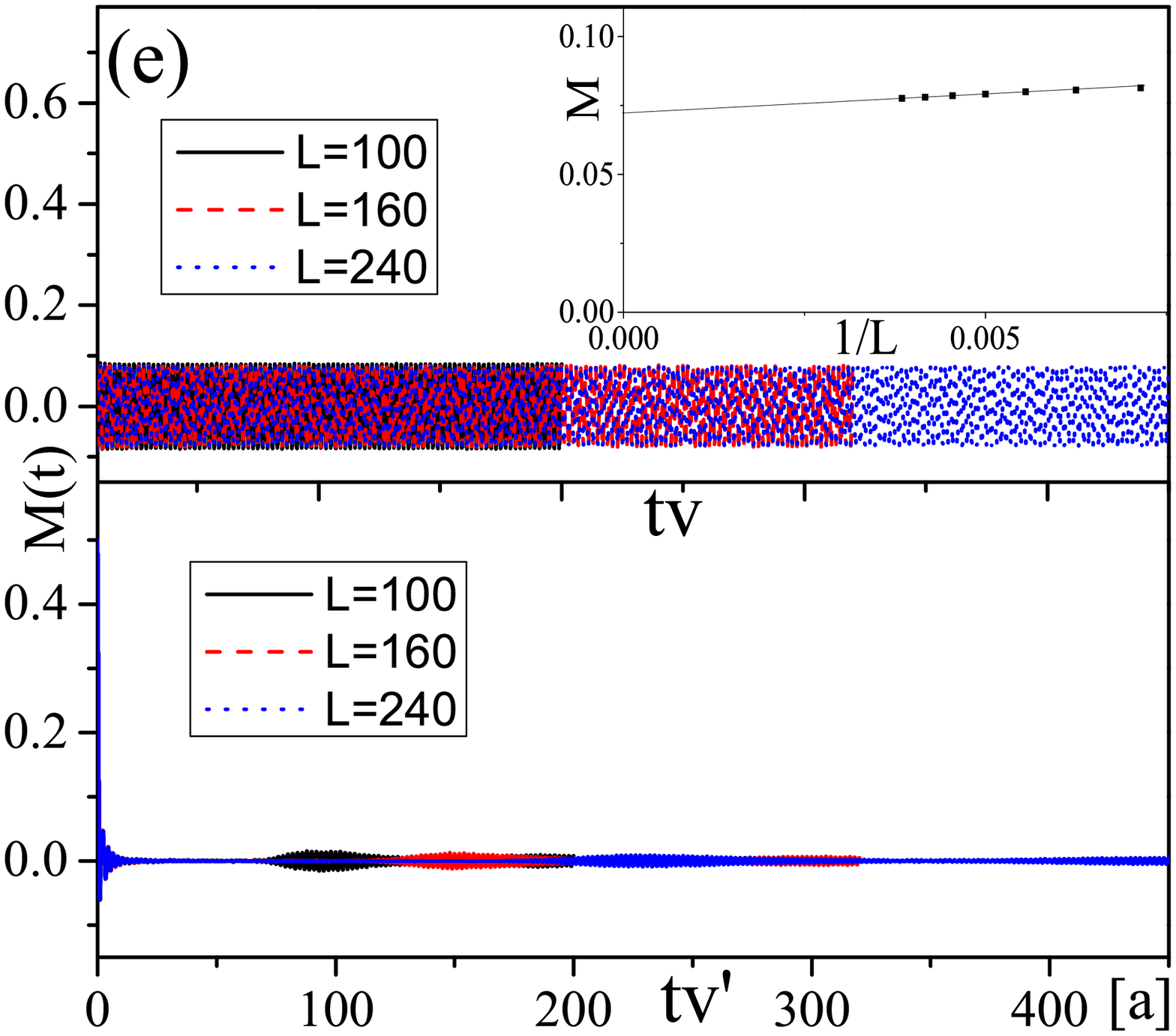}
\includegraphics[width=0.33\linewidth,bb=58 58 609 537]{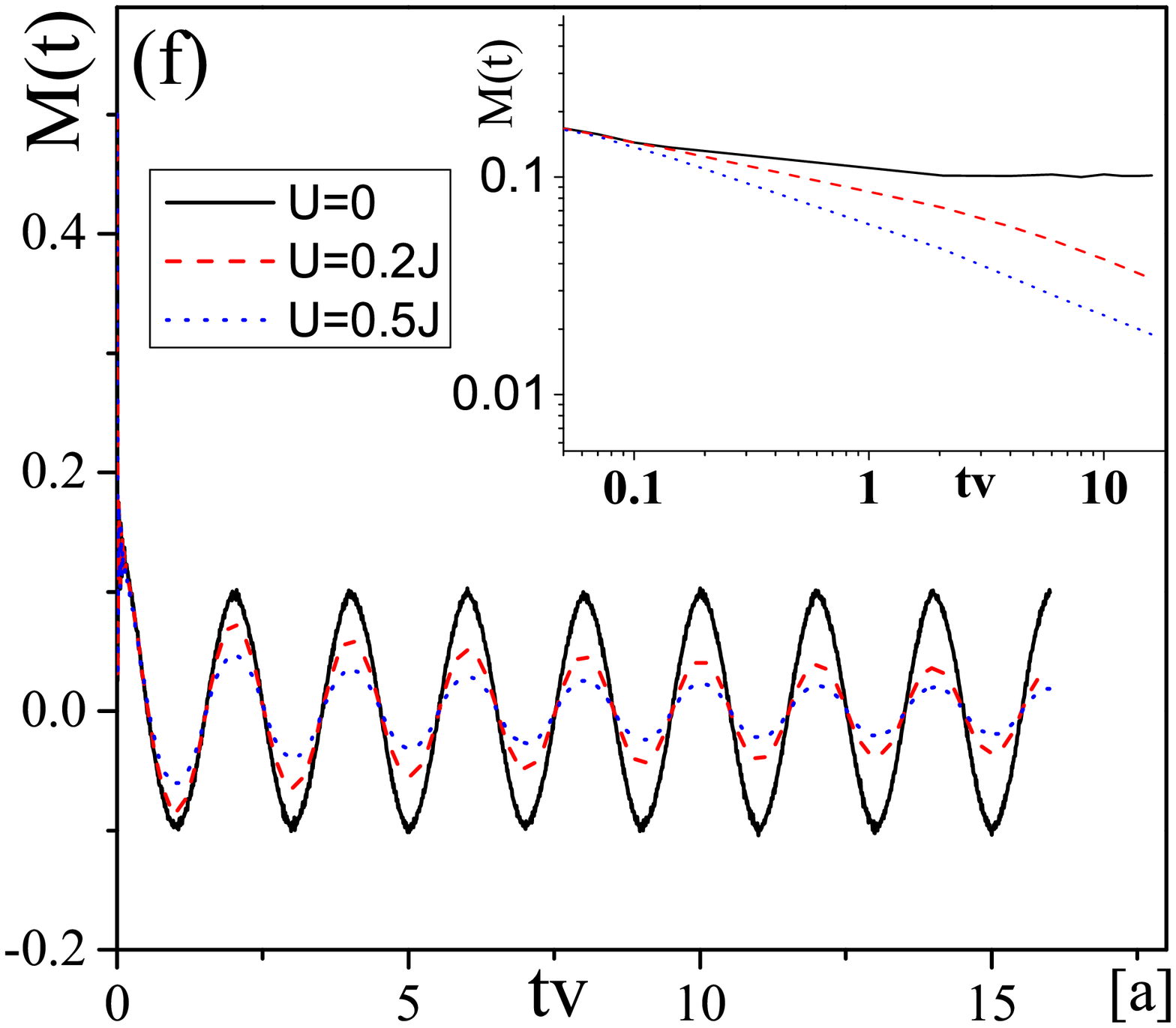}
\caption{(Color online) (a) Density distributions of the wavepacket at different time slices during the single-particle dynamics in the SLP ($v=0.01Ja$) and delocalized phase ($v=Ja$).  The inset magnifies the lower panel. Evolution of the (b) COM and (c) width of the wavepacket under different moving velocities.  Here and hereafter, $tv$ rather than $t$ would be employed to characterize the evolution time, thus allowing the comparison of the dynamics with different $v$ on the same footing.  (d)Comparison between $M(t)$ in SLP ($v=0.01Ja$) and delocalized phase ($v=Ja$) during the non-interacting many-body dynamics beginning from the same ($|1010\cdots\rangle$) initial state. The inset is the envelope of the oscillations in Fig.2 (d) in a log-log plot. (e)Dynamics of M(t) in the non-interacting systems with different system sizes $L$ for the SLP ($v=0.01Ja$ for the upper panel) and delocalized phase ($v'=Ja$ for the lower panel). The inset is a finite size scaling of the amplitudes of the persistent periodic oscillations ($M$) in the SLP ($v=0.01Ja$). (f) Dynamics of M(t) in the small systems with different NN interaction strengths $U$. The inset is the envelope of the oscillations in Fig.2 (f) in a log-log plot. The system sizes are chosen as $L=1000$ for (a)-(c), $L=240$ for (d) and $L=16$ for (f). The discrete time step $\Delta t=0.001J^{-1}$ for all the cases except $v=0.01Ja$ where $\Delta t=0.01J^{-1}$  and $\Delta=4J$ for (a)-(f). A random disorder realization was selected for (a)-(c) and the ensemble averages over 300 disorder realizations were performed for (d)-(f).
} \label{fig:fig2}
\end{figure*}

{\it Single-particle dynamics --} We first consider the single-particle situation, where the initial state was chosen as the ground state of Hamiltonian (\ref{eq:Ham2}) with $t=0$.  Starting from such a spatially localized wavepacket, we study the time evolution of the wavefunction $|\psi(t)\rangle$ by directly solving the time-dependent Schrodinger Equation with different moving velocities. The density distributions $\rho_i(t)=\langle \psi(t)| \hat{n}_i |\psi(t)\rangle$ during the time evolution were plotted in Fig.\ref{fig:fig2} (a), which revealed that  the COM of the wavepacket $X(t)= \sum_i i\rho_i(t)$ could not  catch up with the moving potential at a relatively high velocity  ($X(t)<X(0)+vt$ as shown in Fig.\ref{fig:fig2} b). Consequently, the wavepacket experienced a sequence of rapidly varying disordered potential, which consistently perturbed the wavepacket stochastically as noise, and finally delocalized it.    This delocalization  can be characterized by the width of the wavepacket $w(t)=\sqrt{\sum_i \rho_i(t) [i-X(t)]^2 }$. Fig.\ref{fig:fig2} (c) shows that  $w(t)$ roughly grows linearly after a  long time,  indicating a ballistic transport ($w(t)\sim t$) rather than  diffusion ($w(t)\sim t^{\frac 12}$) reported previously in noisy disordered systems\cite{Gopalakrishnan2017,Cai2020}. This discrepancy exists because  the SSTS (Eq.\ref{eq:symmetry}) resulted in  a strong space-time correlation for the disorder-induced ``noise'', thus stopping it from being  ``white'' noise. Such a ballistic transport resembles the dynamics of the disorder-free case, indicating that a fast-moving disordered potential does not qualitatively change the long-time dynamics of lattice systems, it only renormalizes the hopping amplitude $J$.

 The situation is qualitatively different at a low velocity. Fig.\ref{fig:fig2} (b) shows that the COM adiabatically followed the moving potential ($X(t)=X(0)+vt$), while its width $w(t)$ was bounded even after sufficiently a long time (see Fig.\ref{fig:fig2} c), indicating the absence of diffusion. The uniform motion of  COM and the absence of the diffusion of the wavepacket were the main features of SLP at the single-particle level. The existence of the SLP indicates the validity of the adiabatic theorem in our model\cite{Supplementary}. The localization/delocaliztion of a wavepacket does not only depend on the moving velocity, but also on the initial state energy. For instance, if we begin from another eigenstate of $\hat{H}(t=0)$ close to the band edge, the physics is similar as analyzed above, while for an initial state in the band center, even an infinitesimal velocity would be sufficient to delocalize the wavepacket. Consequently, for a given velocity, a critical initial energy exists to separate the sliding localized and delocalized states. The existence of such a critical initial energy, is crucial for our subsequential discussion regarding the many-body cases with and without interactions.

It is helpful to compare our results with the wavepacket dynamics under a suddenly moving single trap. For instance, the dynamics of a bound state in the presence of a uniformly moving attractive $\delta$ potential (with a velocity $v$) can be solved exactly\cite{Granot2009}, which showed that the wavepacket will be split into three parts with different velocities: one adiabatically follows the moving trap at a velocity $v$, while the other two escape from the trap (with a velocity $0$ and $2v$ respectively), and there is no dynamics phase transition. In our model, the two wavepackets escaping from the original trap will be localized by the disordered potential in the SLP, thus the wavepacket will adiabatically follow the moving potential.

 \begin{figure}[htb]
\includegraphics[width=0.99\linewidth,bb=107 23 767 360]{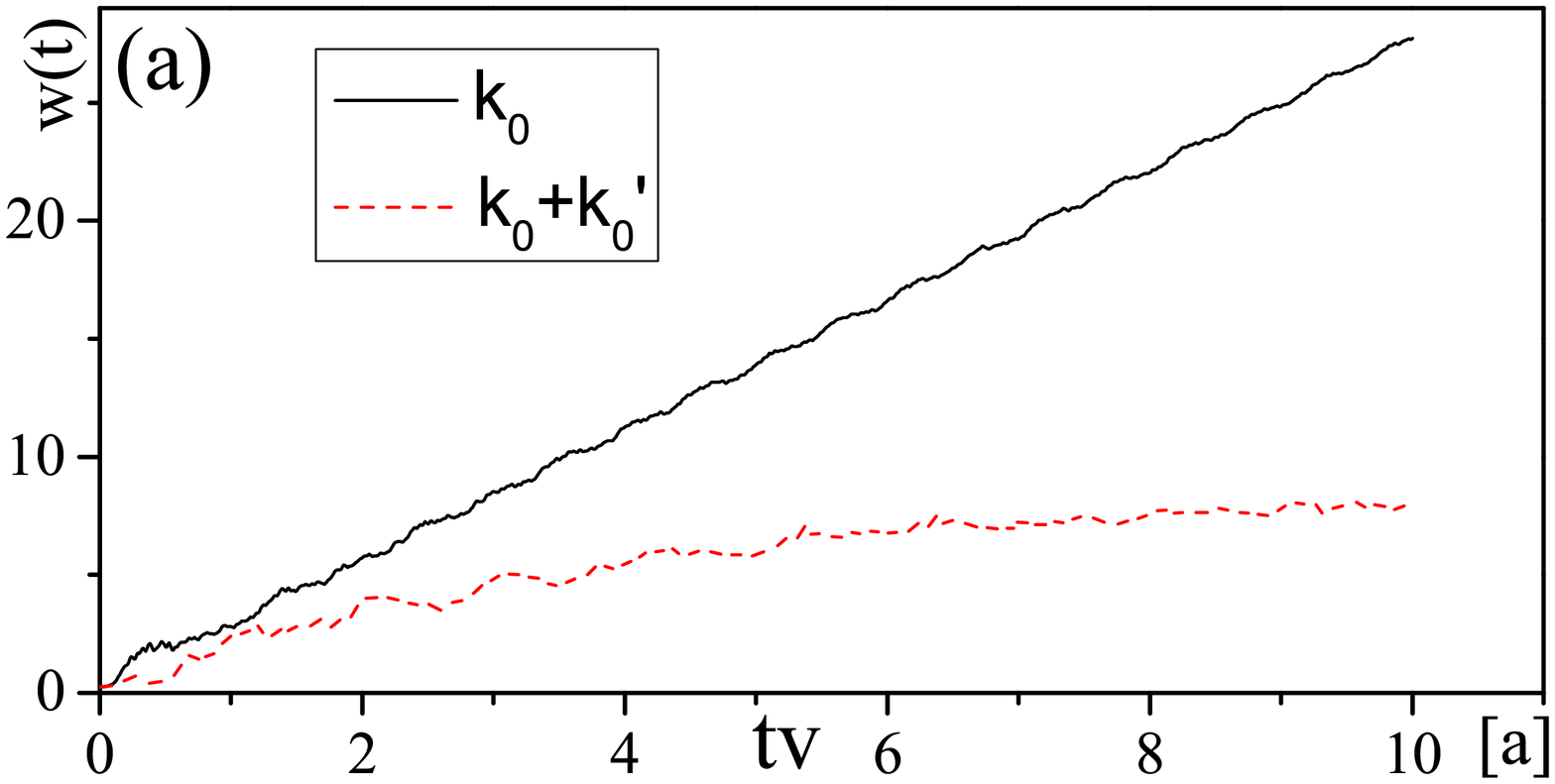}
\includegraphics[width=0.9\linewidth]{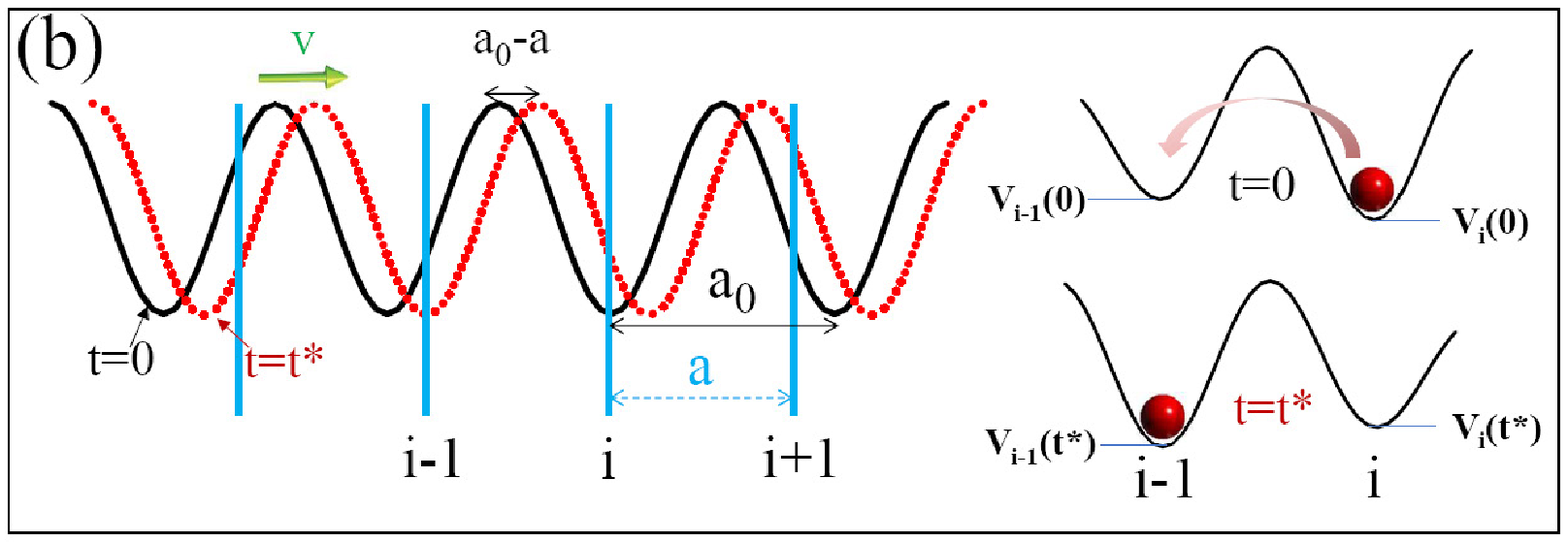}
\caption{(Color online) (a) Comparison between the dynamics of the width of wavepacket in the presence of a single moving quasi-periodic potential $V_{k_0}(x,t)=\Delta \cos 2\pi k_0(x-vt)$  with and a superposition of two quasi-periodic potentials $V'_{k_0}(x,t)=\frac\Delta 2 [\cos 2\pi k_0(x-vt)+\cos 2\pi k'_0(x-vt)]$. The parameters are chosen as $\Delta=4J$, $L=1000$, $v=0.01Ja$, $k_0=\frac{\sqrt{5}-1}2$ and  $k'_0=\frac{\sqrt{6}-1}2$. (b)Sketch of the nearest-neighboring tunneling of the wavepacket induced by a moving qPP during the period $[0,t^*]$ with $t^*=\frac{a_0-a}v$.}\label{fig:fig3}
\end{figure}

{\it Many-particle dynamics without interaction --} One of the valuable features of localization is the memory effect of the initial-state information, which will be partially preserved  during  time evolution.  For instance, if we begin from a half-filled charge-density-wave(CDW) state $|1010\cdots\rangle$, the density imbalance between two sublattices of the 1D lattice $M(t)=\frac 1L\sum_i (-1)^i\rho_i(t)$  in the initial state would be memorized ($M(t\rightarrow \infty)\neq 0$) in localized phases\cite{Schreiber2015}; however, in  delocalized phase, it would be washed out after extended period ($M(t\rightarrow \infty)\rightarrow 0$). Thus, to characterize the feature of SLP from the perspective of many-body physics, we selected the $|1010\cdots\rangle$ initial state  and studied the time evolution under the noninteracting Hamiltonian (\ref{eq:Ham2}) with different $v$  by solving the equation of motion of the equal-time single-particle Green's function $G_{ij}(t)=\langle \psi(t)| C_i^\dag C_j |\psi(t)\rangle$, from which all the physical quantities at any given time, including $M(t)$, were derived.

Without interaction, the dynamics of the many-body (fermionic) systems can be understood as the collective behavior of different single-particle states, each of which evolves independently. At a low velocity, the system adiabatically followed the moving potential, so is the CDW state. A CDW state moving at a constant velocity can be characterized by the persistent periodic oscillation of $M(t)$ (Fig.\ref{fig:fig2} d). Notably the oscillation amplitude is smaller than that in the initial state ($M(0)=0.5$), which can be explained by the aforementioned delocalized single-particle states that barely contributed to the sliding CDW order. $M(t)$ in systems with different system sizes were plotted in Fig.\ref{fig:fig2} (e). Therein, a finite-size scaling (see the inset and \cite{Supplementary}:) shows that such periodic oscillations in SLP persisted in the thermodynamic limit. Although such a sliding CDW state resembles the ``moving solid''  proposed in the steady states of  CDW systems driven by applied electric fields\cite{Koshelev1994,Marley1995,Balents1995}, there is significant difference: the CDW order in the ``moving solid'' is induced by interactions, while that in our study originated from the initial state memory.   At a high velocity, the long-time dynamics of $M(t)$ noticeably exhibited an algebraic decay $\sim t^{-\alpha}$ (accomplished by an oscillation) with a non-universal exponent $\alpha$ depending on $v$.

{\it Many-particle dynamics with interaction --}  To examine the  effect of interaction on the SLP in our model, the NN  interactions between fermions were introduced as follows:
\begin{equation}
\hat{H}_I(t)=\sum_i [-J(C_i^\dag C_{i+1}+h.c)+V_i(t) \hat{n}_i + U \hat{n}_i \hat{n}_{i+1}] \label{eq:Ham3}
\end{equation}
where U is the strength of the NN interaction.  Employing the exact diagonalization method, the time evolution of $M(t)$ was studied  beginning from the $|1010\cdots\rangle$ state under the Hamiltonian (\ref{eq:Ham3}) for a relatively small system ($L=16$) and compared with the non-interacting case.  Fig.\ref{fig:fig2} (f) shows that even a weak interaction  in the SLP ($v=0.01J$) can delocalize the system and lead to an algebraic decay of $M(t)$. This is strikingly different from the static disorder-induced localization, which is generally robust against weak interaction. Such an interaction-induced delocalization can also be explained by the aforementioned critical initial energy, above which the single-particle states are delocalized. Generally, the interaction would induce scatterings between the different single-particle states, thus mix the localized and delocalized states. Therefore, those delocalized states above the critical initial energy act as a bath that is coupled to the localized states, and delocalize the whole system.


{\it Discussion--} Under Fourier transformation  $\tilde{V}(x,t)=\frac{1}{\sqrt{L}}\sum_k V_k e^{ik(x-vt)}$, the moving disorder potential turns to a superposition of a set of moving periodic potentials with different wavelengths, which are typically incommensurate with the lattice constant $a$. Therefore, to understand the SLP, one should first focus on the dynamics of a particle in the presence of a moving quasi-period potential (qPP) with a period incommensurate with $a$. Regarding the static case ($v=0$), it is known that a qPP with $\Delta>2J$ would also result in a  localization that is similar to the Anderson localization in the disordered system\cite{Aubry1980}. Thus, their dynamical behaviors in the presence of the moving potential could be expected to be similar. However, Fig.\ref{fig:fig3} (a) shows this is not the case. In the presence of a slowly moving qPP $V_{k_0}(x,t)=\Delta \cos 2\pi k_0(x-vt)$ with $v=0.01J$, starting from the initial state as the ground state of $V_i=V_{k_0}(ia,0)$, the variance of the wavepacket w(t) is plotted in Fig.\ref{fig:fig3} (a) (the solid black curve), which grows linearly with time, indicating a ballistic transport behavior similar with that of disorder-free case.

The delocalization of a particle under a slowly moving qPP can be understood via the Landau-Zener tunneling. Considering a qPP with a period $a_0$ incommensurate but slightly larger than $a$, and further assuming that initially the minima of the qPP and lattice potential coincide at site i, we choose the initial state as the ground state at $t=0$, which spatially localized around site i. Since $a_0$ is slightly larger than $a$, the potential energies on site $i$ and $i\pm 1$ at $t=0$ are close with each other but separated from those on other sites. We focus on site $i$ and $i-1$, whose potential energies vary with the movement of qPP. As shown in Fig.\ref{fig:fig3} b, at time $t^*=\frac{a_0-a}v$, the energy minimum  has been  shifted from site $i$ to $i-1$ ($V_{i-1}(t^*)=V_i(0)$), indicating an energy level crossing. Considering the particle tunneling between site $i$ and $i-1$, this physics resembles the Landau-Zener tunneling: for a slow moving velocity $v\ll J$, the wavepacket will follow the energy minimum thus tunnel from site $i$ to site $i-1$ at $t=t^*$. With  further movement of qPP, the wavepacket will keep tunneling, thus give rise to a ballistic transport with a renormalized tunneling rate depending on $t^*$\cite{Supplementary}.

This picture is helpful for us to understand the SLP in the moving disordered potential, which can be considered as a superposition of  moving qPPs with different $a_0$. For a single qPP, a wavepacket tunneling from site $i$ to $i-1$ carries a phase  depending on the tunneling time $t^*$, which in turn is determined by $a_0$ of the qPP. Therefore, in the presence of many qPPs with random $a_0$, the wavepacktes carrying different phases will interference with each other at site $i-1$, thus suppress the effective tunneling and lead to localization. This point can be numerically verified by comparing the dynamics of $w(t)$ with a single qPP and more than one qPP. As shown in Fig.\ref{fig:fig3} (a), in the presence of two qPPs with different periods, $w(t)$ significantly deviates from the linear growth, indicates that the effective tunneling is strongly suppressed.

{\it Experimental realization and detection-- } The proposed model can be experimentally simulated by loading  atoms  into a quasi-1D optical lattice, in which the disordered and quasi-periodic potentials can be introduced in controllable ways: the former was realized by  by implementing a speckle potential generated by a laser beam passing through a diffusion plate\cite{Clement2006,Billy2008}, while the latter could be introduced by imposing an additional optical lattice whose period is approximately incommensurate with that of the original one\cite{Roati2008}.  In both cases, the moving potentials could be realized by shifting either the diffusion plate or the phase of the additional lasers at a constant velocity. For an optical lattice with  $a=512nm$ and $J\simeq 400Hz$, one can estimate that the realistic moving velocity  corresponds to the typical parameter regime discussed above is in the order of a magnitude of $10\mu m/s$.   Regarding the detections, the density imbalance $M(t)$ can be  be directly or indirectly measured via the superlattice band-mapping technique\cite{Trotzky2012}, or heterodyne detection method respectively\cite{Landig2016,Hruby2018}

{\it Conclusion and outlook -- } In summary, the dynamics of quantum systems with moving disordered potentials were studied, and a sliding localized phase, which availed a new perspective for studying of driven-disordered systems, was clarified. This result does not only apply to matter waves in quantum systems, but also to a wide range of classical wave phenomena in disordered media, such as light\cite{Wiersma1997}, sound\cite{Weaver1990} and microwaves\cite{Dalichaouch1991}.  Future developments will include generalization/extension of our findings to higher-dimensional systems, in which the motion of  disordered potentials along a certain dimension might produce intriguing anisotropic localized phases, in which the matter wave is delocalized along the driving direction, but still  localized along other directions that are perpendicular to it. A more general question is whether there exists an effective  time-independent description analog to the Floquet Hamiltonian  for this type of driven systems with intertwined space-time symmetry. If there exists any, what are the  ``local'' integrals of motions that account for the initial state memory in the sliding localized phase? Finally, even though we focused on localization physics, the proposed driving protocol can be studied in a broader context, especially in the interacting quantum systems where the interplay between the spontaneous symmetry breaking and the moving disorder might give rise to nontrivial phenomena, {\it e.g.,} a generalization of the Imry-Ma arguments\cite{Imry1975} for the robustness of symmetry breaking to such a moving disorder.

{\it Acknowledgments}.---This work is supported by the National Key Research and Development Program of China (Grant No. 2020YFA0309000), NSFC of  China (Grant No.12174251), Natural Science Foundation of Shanghai (Grant No.22ZR142830),  Shanghai Municipal Science and Technology Major Project (Grant No.2019SHZDZX01).


\newpage
\begin{center}
\end{center} 

\textbf{\Large{Supplementary material for ``Matter waves in lattice systems with moving disorder''}}

{\small \setcounter{tocdepth}{1} \tableofcontents }

\section{Details of the numerical simulations}
In this section, we check the dependence of the sliding localized phase on different disorder realizations, system sizes, and spline interpolations.

\subsection{Disorder realizations}
\begin{figure}[htb]
\includegraphics[width=0.99\linewidth,bb=64 56 1100 560]{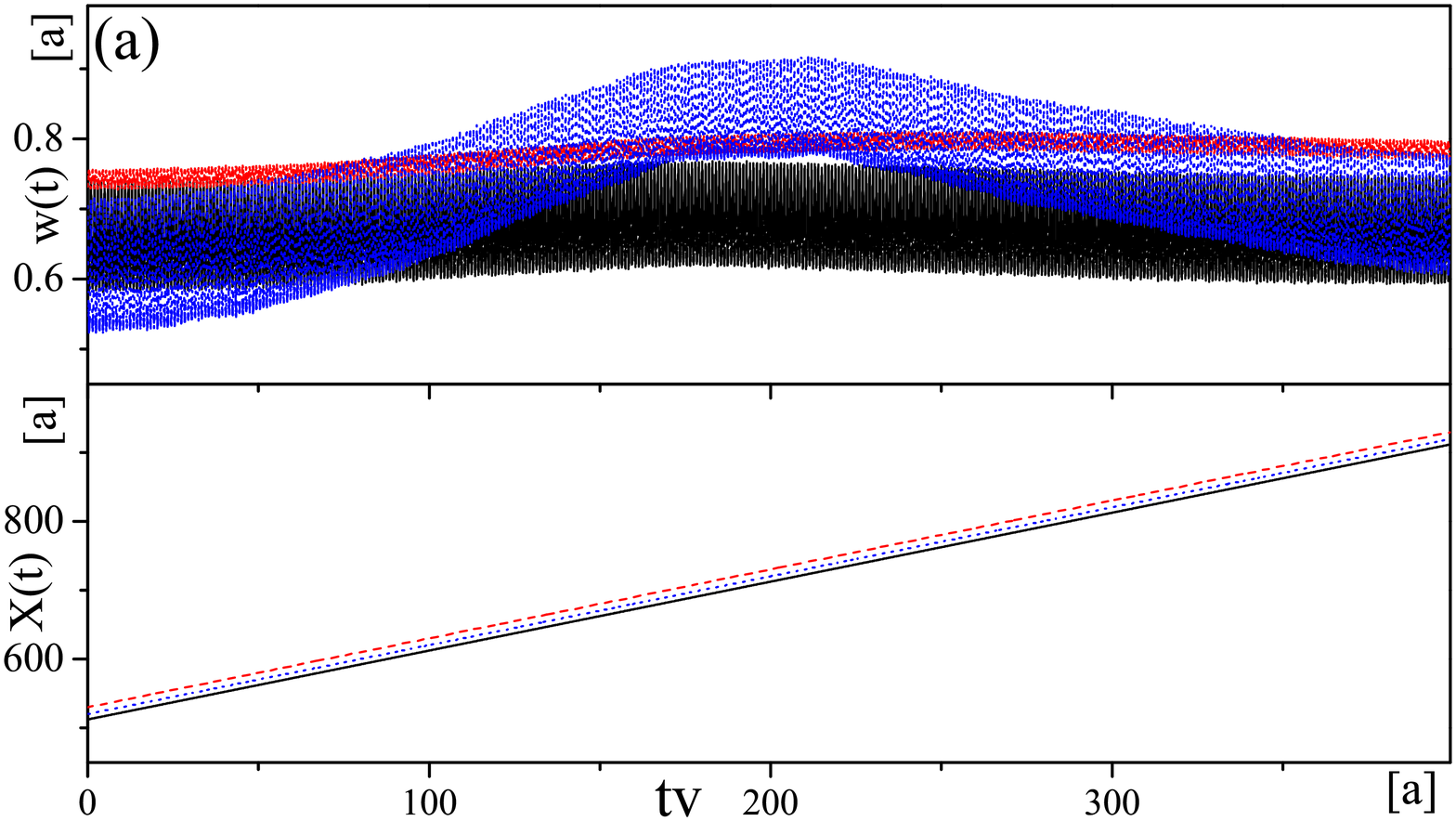}
\includegraphics[width=0.99\linewidth,bb=64 56 1100 560]{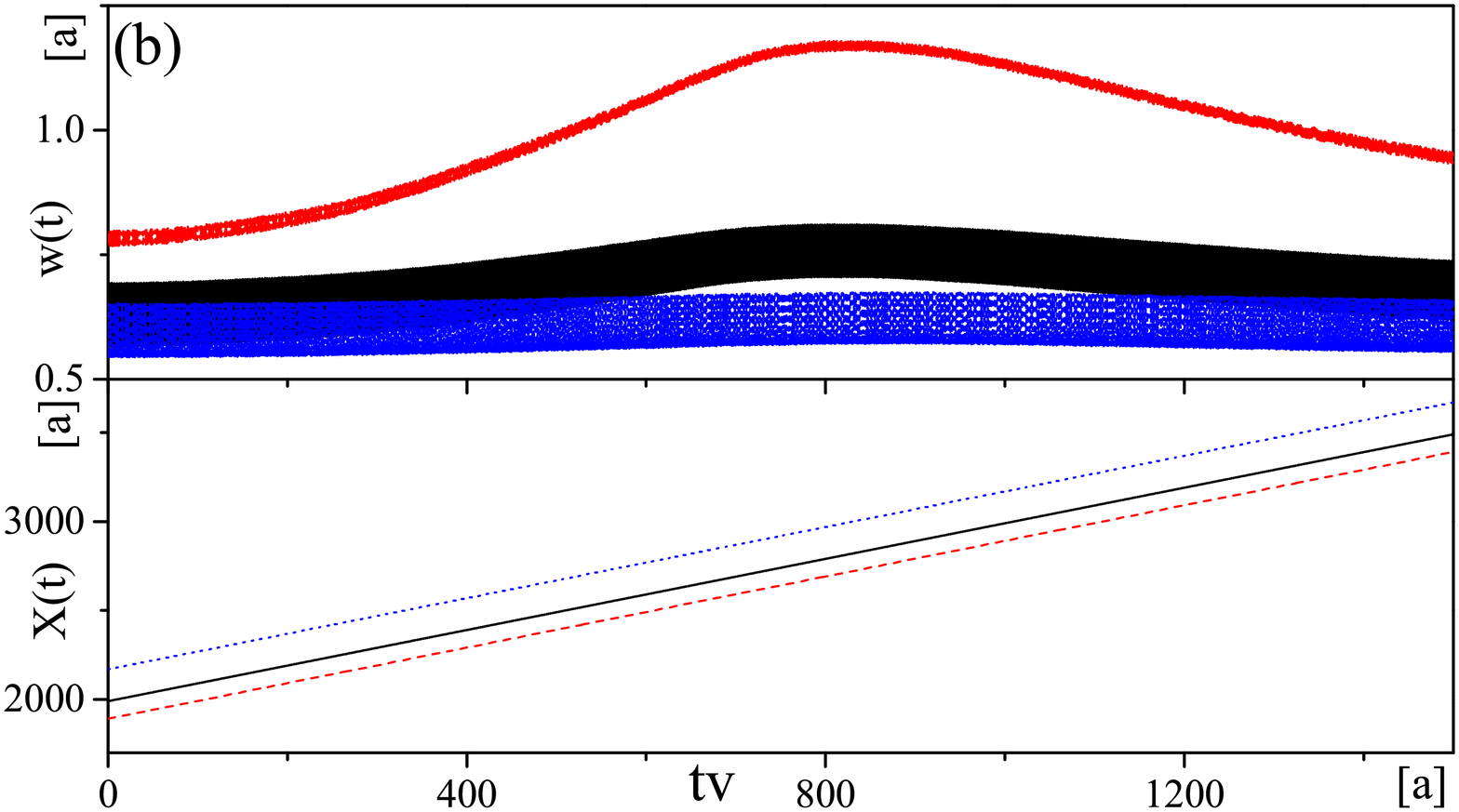}
\caption{(Color online) The time evolution of COM (upper panel) and width (lower panel) of the wave packets under three different disorder realizations with system size  (a) L=1000 and (b) L=4000, $v=0.01Ja$ and $\Delta=4J$. }\label{fig:SM1}
\end{figure}

In the maintext, the single-particle dynamics is calculated under a  given set of randomly chosen disorder realization $\{V_i(t=0)\}$, it is thus important to check the our results do not crucially depend on the choices of the disorder realizations. To this end, we calculate the dynamics of the center of mass (COM) and width of the wavepackets for different disorder realizations $\{V_i(t=0)\}$ starting from the initial state as the ground state of the corresponding Hamiltonian. Without losing generality, we choose those disorder realizations whose ground state locates close to the center of the 1D lattice to avoid the boundary effect during the evolution as possible as we can. As shown in Fig.\ref{fig:SM1} (a), the long-time dynamics of the wavepackets with different disorder realizations are qualitatively identical: at a low velocity ($v=0.01Ja$), the COMs of the wavepackets under different disordered potential adiabatically follow the moving potential: $X(t)=X(0)+vt$, while in all these cases, their widths keep oscillating within a finite regime, which indicates an absence of diffusion. Therefore, the existence of the sliding localized phases doesn't depend on the choices of the disorder realization.

\begin{figure}[htb]
\includegraphics[width=0.9\linewidth,bb=129 56 900 560]{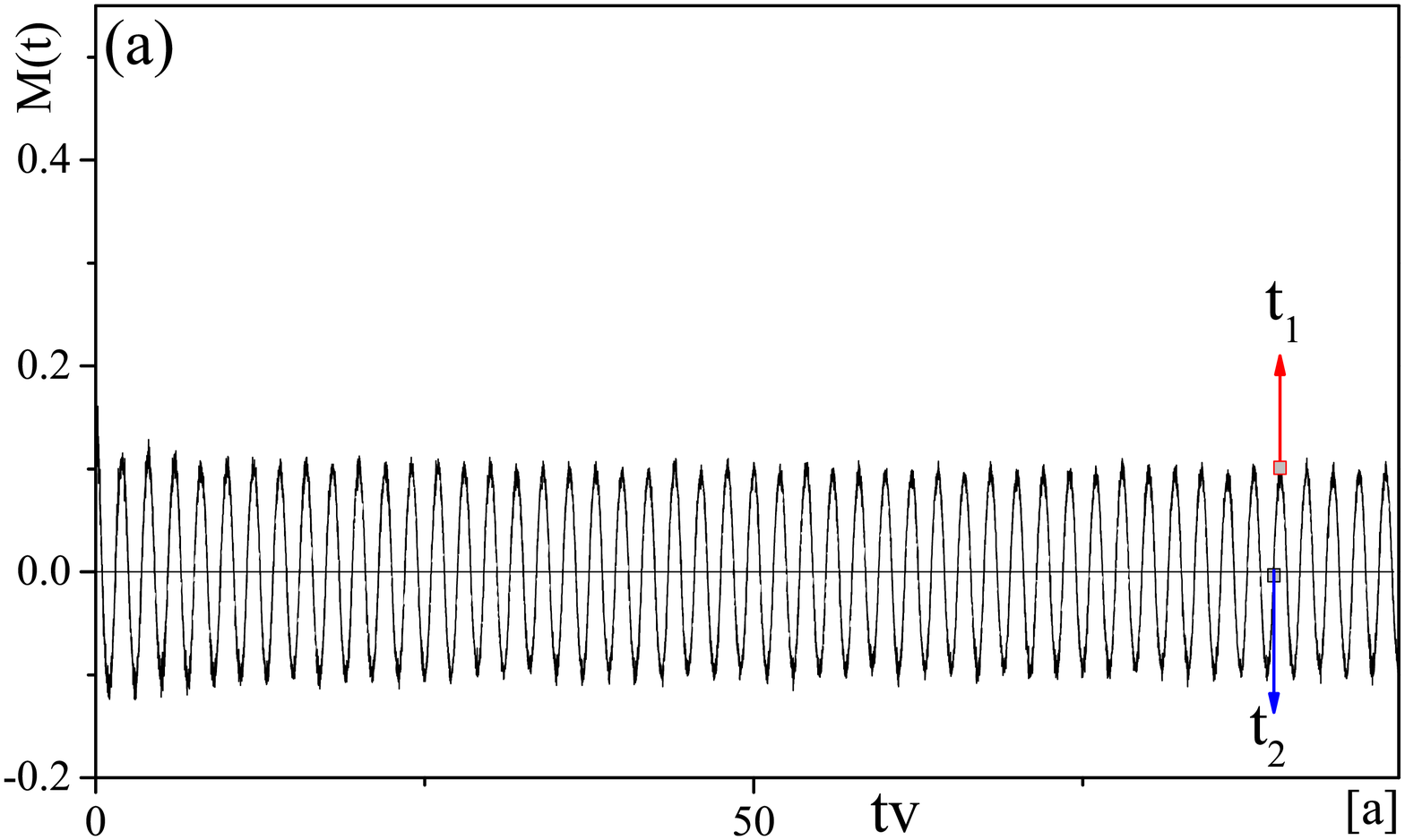}
\includegraphics[width=0.9\linewidth,bb=129 56 900 560]{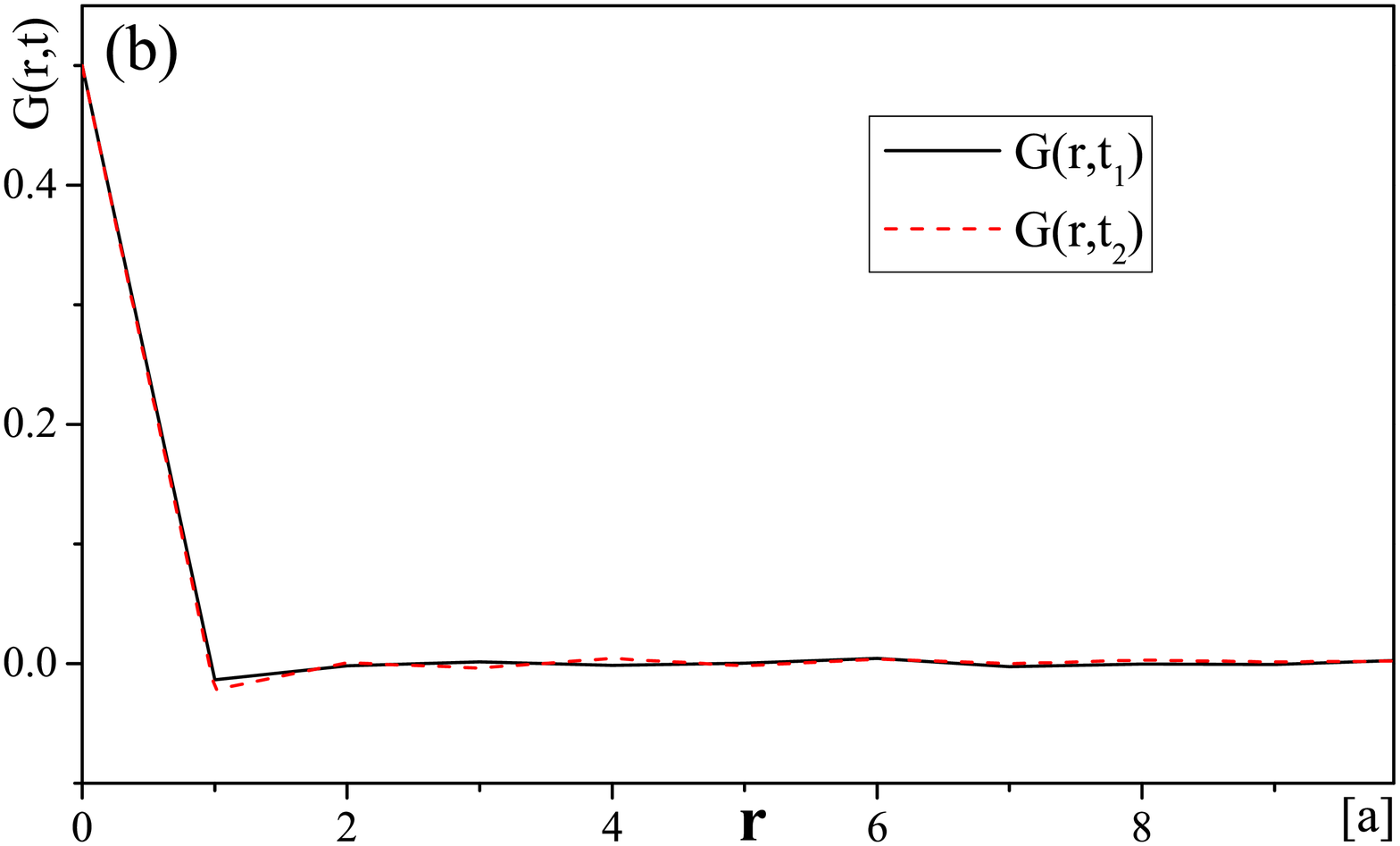}
\caption{(Color online) (a) The dynamics of M(t) for a non-interacting many-body system starting from the $|1010\cdots\rangle$ with parameters $L=240$, $\Delta=4J$, $v=0.01Ja$. (b) The equal-time single particle Green's function at two different time slices ($t_1$ and $t_2$), which  respectively corresponds to a wave peak and node of $M(t)$ as shown in (a).  Ensemble average  over 300 disorder realizations are performed in (a) and (b).}\label{fig:SM2}
\end{figure}

\begin{figure}[htb]
\includegraphics[width=0.99\linewidth,bb=50 51 750 422]{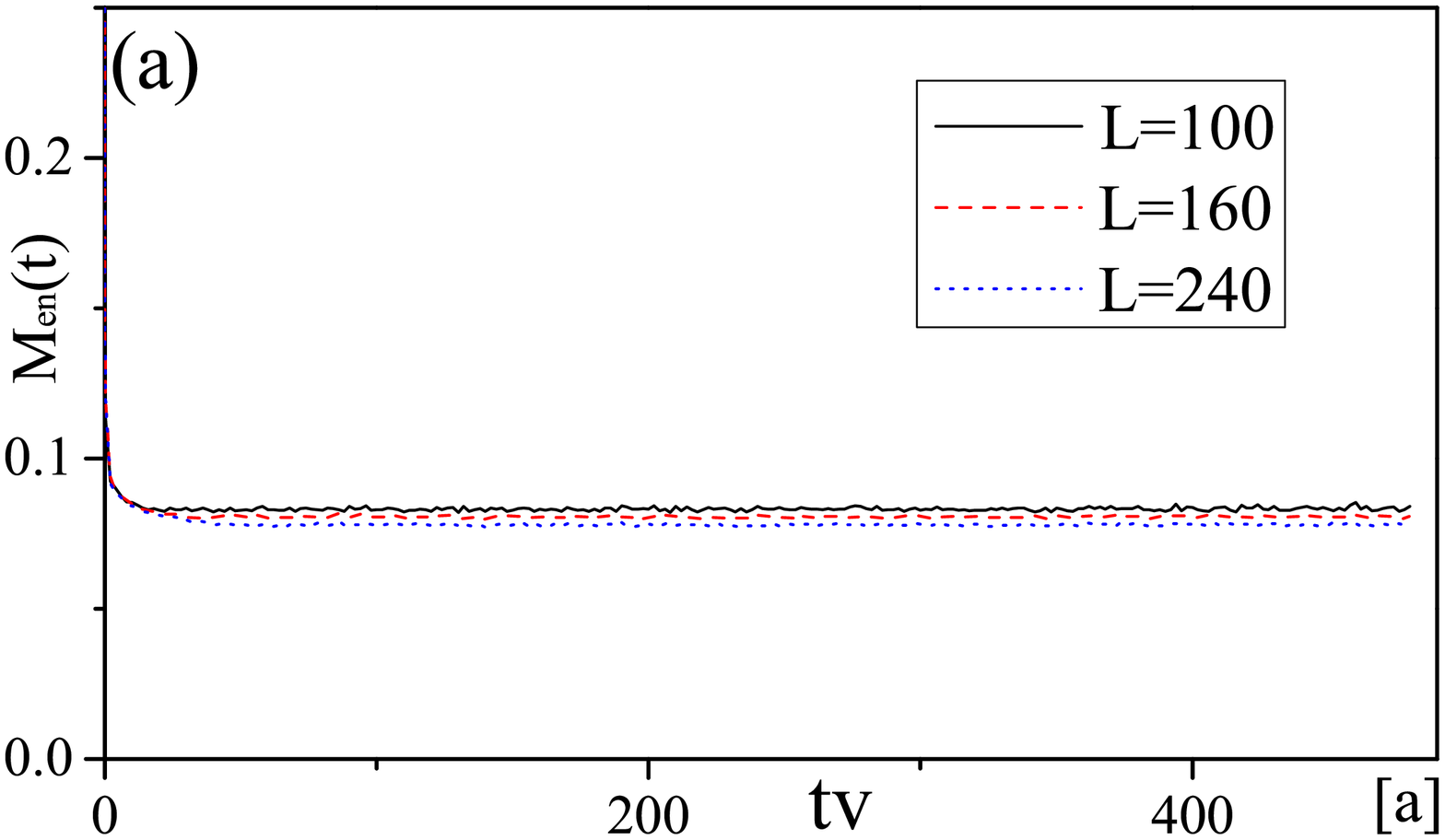}
\includegraphics[width=0.99\linewidth,bb=50 51 750 422]{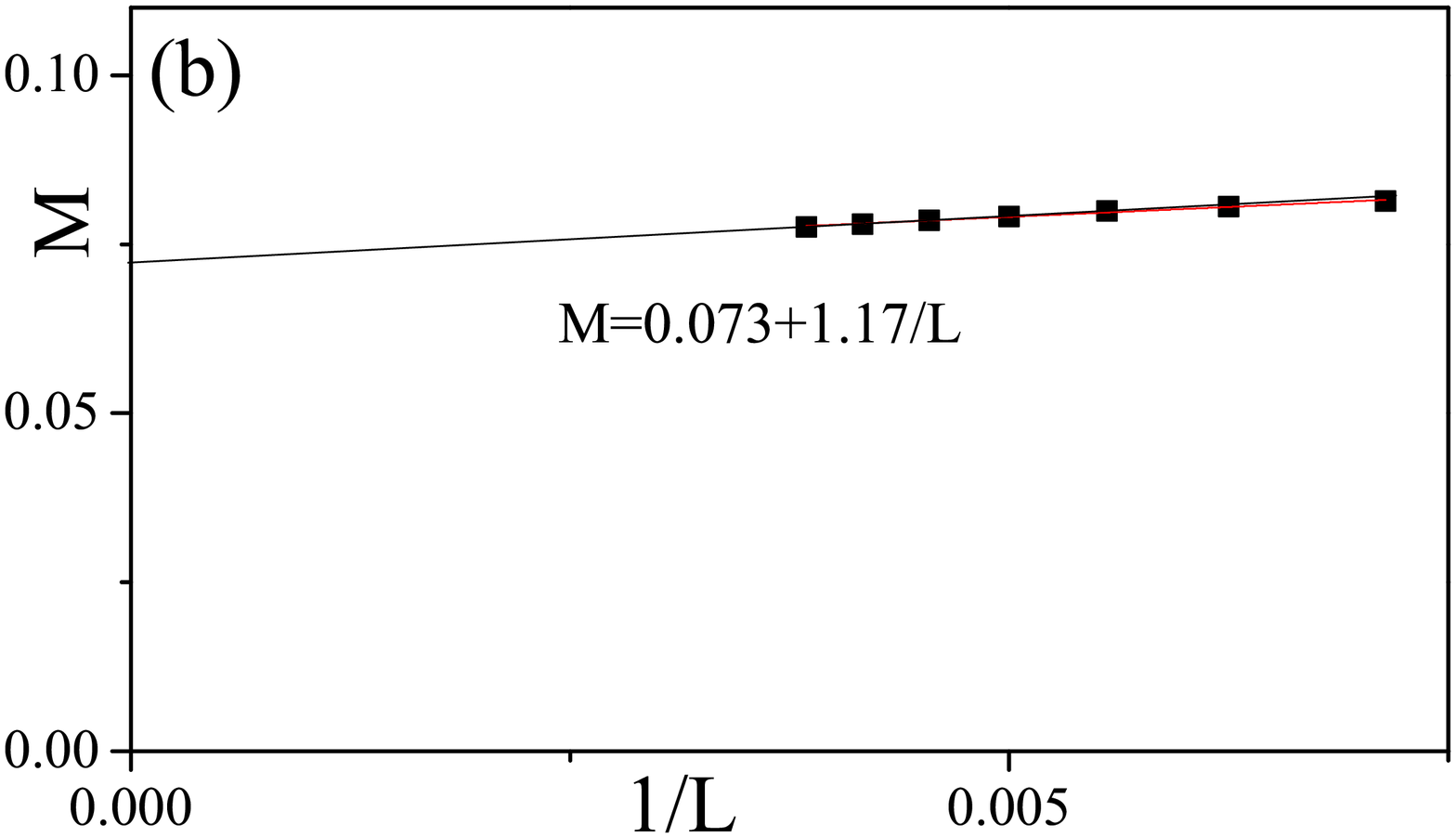}
\caption{(Color online)(a) The envelop of M(t)  for different system sizes in the many-body dynamics of SLP.   (b) Finite-size scaling of the saturated  amplitude of the oscillation M.  $\Delta=4J$ and $v=0.01Ja$ for (a) and (b). 300 disordered realizations are performed for each system size.}\label{fig:finitesize}
\end{figure}

\begin{figure}[htb]
\includegraphics[width=0.9\linewidth,bb=129 56 900 560]{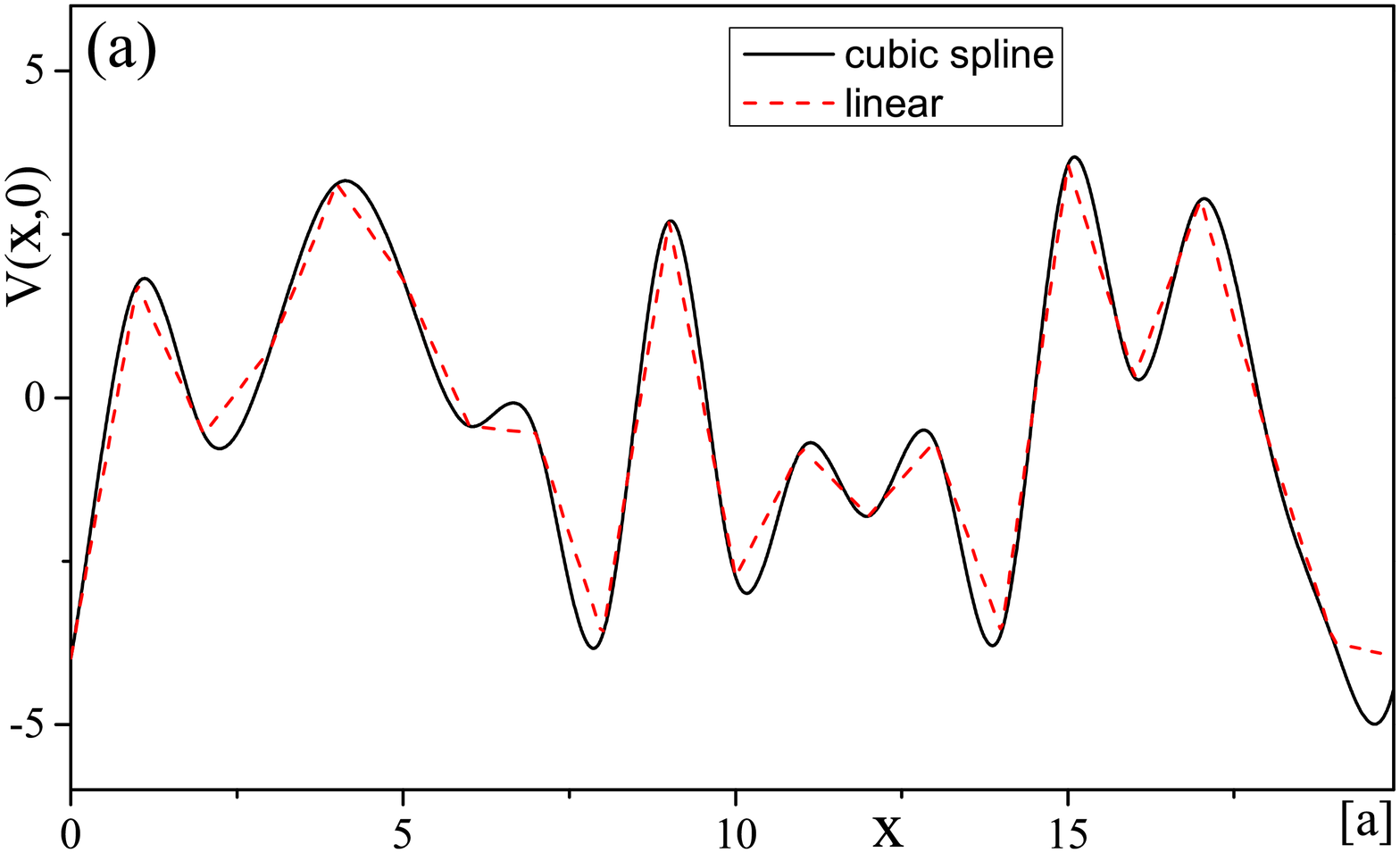}
\includegraphics[width=0.99\linewidth,bb=43 56 800 560]{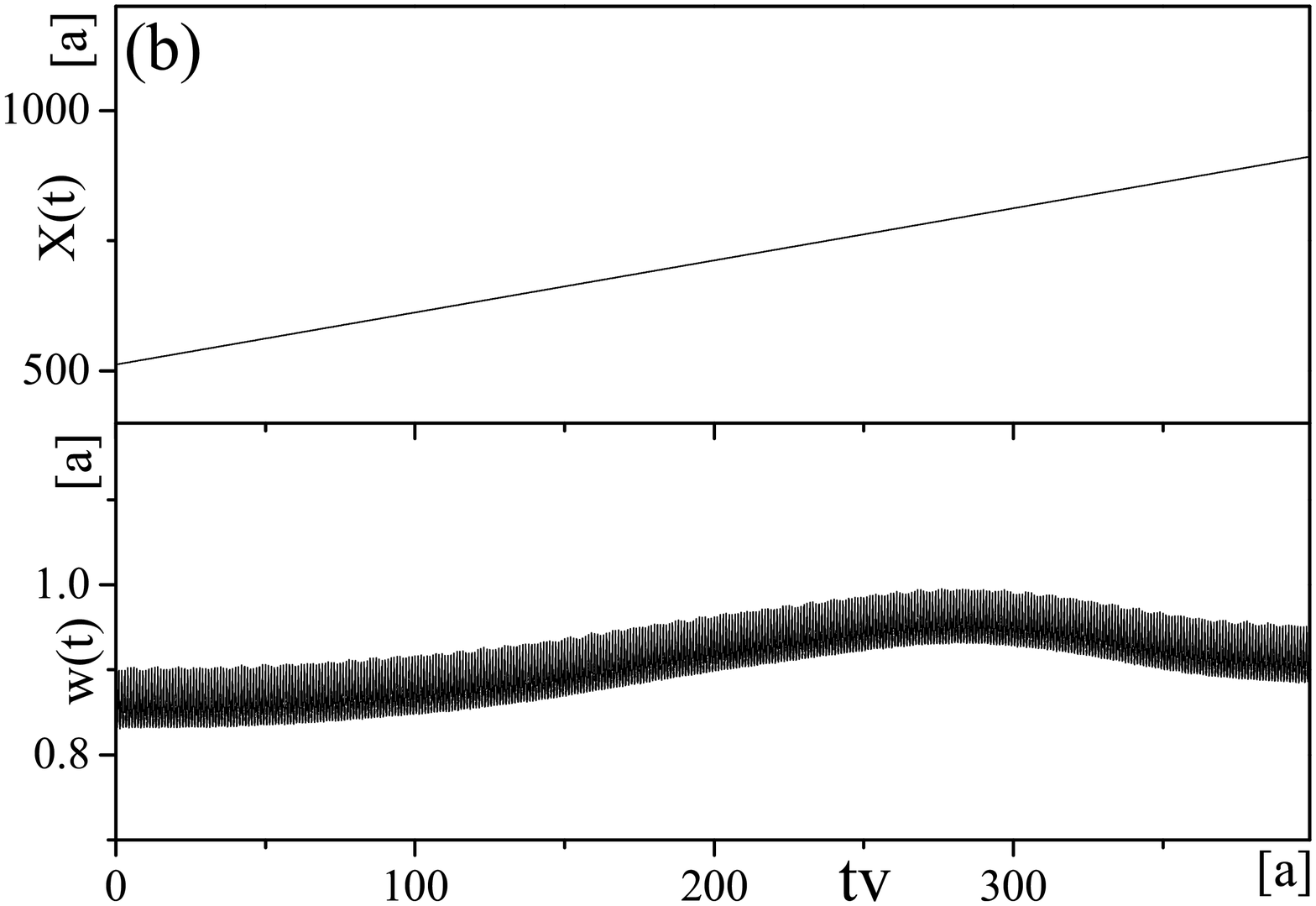}
\caption{(Color online)(a) A comparison between the cubic spline interpolation and the linear interpolation from a same set of random $\{V_i(t=0)\}$ (b) The dynamics of COM (upper panel) and width (lower panel) of the wavepacket under the moving potential $\tilde{V}(x,t)$ derived from the linear interpolation with parameters $L=1000$, $\Delta=4J$, $v=0.01Ja$.}\label{fig:SM3}
\end{figure}

\subsection{System size dependence and localization length}

The single-particle simulation in the main text is performed on a 1D lattice with finite system site(L=1000), which also limits the maximum simulation time ($t_{max}\sim  La/v$), after which the wave packet will be bounced by the boundary of the 1D lattice under the open boundary condition. For a 1D system with periodic boundary condition, in principle, the COM of a wavepacket located on a ring is ill defined.  To avoid the boundary/finitesize effect, we choose those initial states localized close to the middle of the 1D lattice ($X(t=0)\simeq L/2$), and the typical simulation time $T<t_{max}/2$.

It is known that sometimes the localization phenomena may be sensitive to the system size, especially for the weak disordered cases where the localization length is long enough to be compatible to the system size.  To make sure that the system size ($L=1000$) and simulation time chosen in the main text are  long enough to capture the physics of the infinite long-time dynamics of a system in the thermodynamic limit,  here we study a larger system ($L=4000$), which also allows us to simulate the dynamics with a longer time. The dynamics COM and the width of the wave packets in a system with $L=4000$, $v=0.01Ja$, but different disorder realizations are shown in Fig.\ref{fig:SM1} (b), from which we can see that there is no qualitatively difference between the results of $L=4000$ and $L=1000$ shown in Fig.\ref{fig:SM1} (a).

In the single particle dynamics, it is difficult to derive the localization length due to the intrinsic non-equilibrium nature of our model, which is different from the conventional Anderson localization or periodically driven systems  where  the localization length can be derived from the eigenstates of the single-particle (Floquet) Hamiltonian. In our model, there is no effective Hamiltonian description, thus  the eigenstate analysis is invalid. However, as shown in the main text, the many-body dynamics of the sliding localized phase starting from the $|1010\cdots\rangle$ state will approach to an asymptotic periodically oscillating dynamics, which allows us to study the equal-time single particle correlation function $G(r,t)=\langle \psi(t)|\frac 1L\sum_i C_i^\dag C_{i+r}|\psi(t)\rangle$ in this asymptotic regime. We calculate $G(r,t)$ at two different time slices ($t_1$ and $t_2$ as shown in Fig.\ref{fig:SM2} a), which respectively correspond to a wave peak and a wave node in the asymptotic oscillation regime.  As shown in Fig.\ref{fig:SM2} (b), both  $G(r,t_1)$ and $G(r,t_2)$ rapidly (exponentially) decay in distance. By making an analogue to the Anderson localization, such a exponential decay of the single-particle Green's function enable us to estimate the localization length (a few lattice constants), which is significantly smaller than the typical system sizes in our simulation (hundreds of lattice constant).

\begin{figure}[htb]
\includegraphics[width=0.99\linewidth,bb=50 51 760 560]{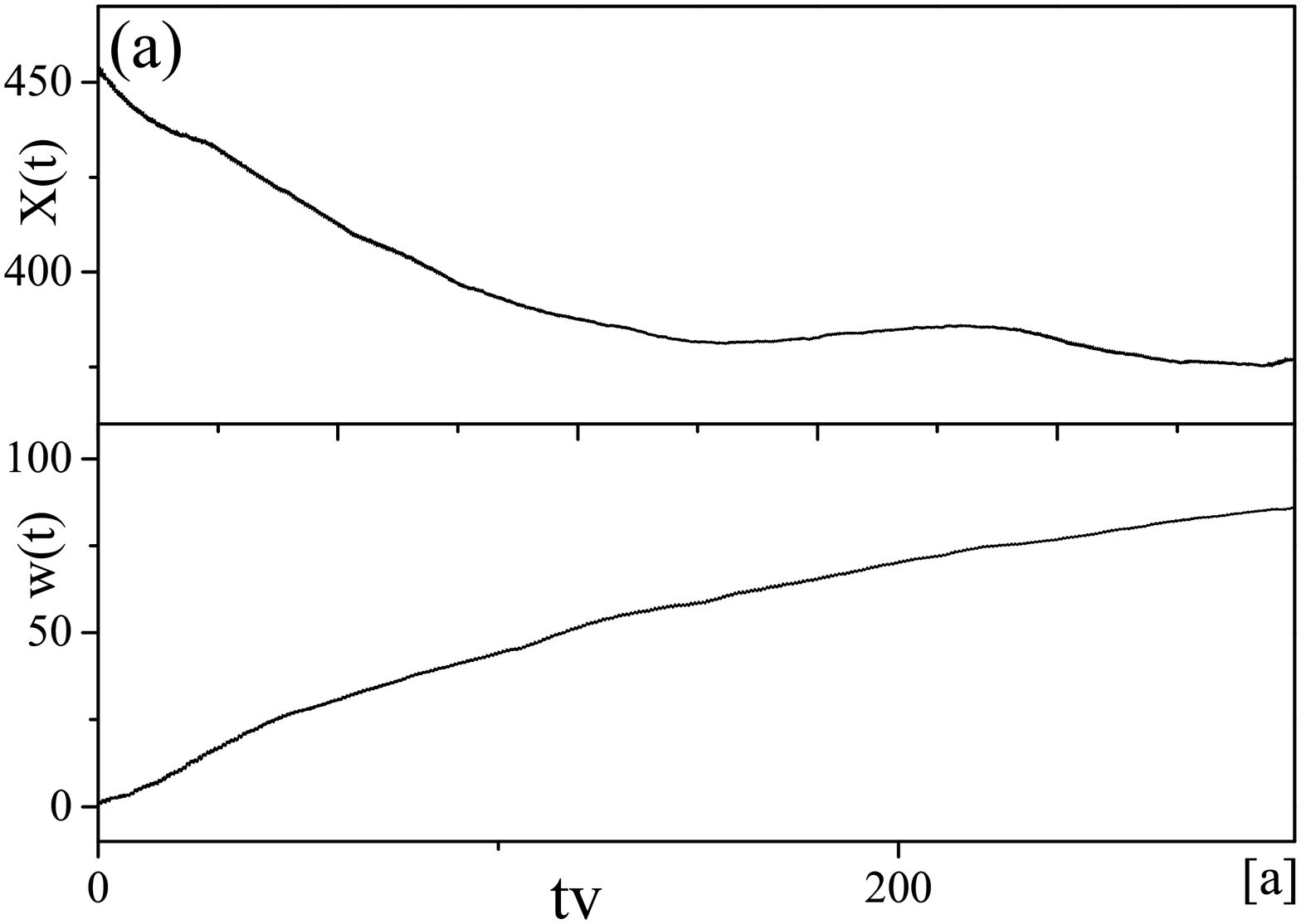}
\includegraphics[width=0.99\linewidth,bb=50 51 760 560]{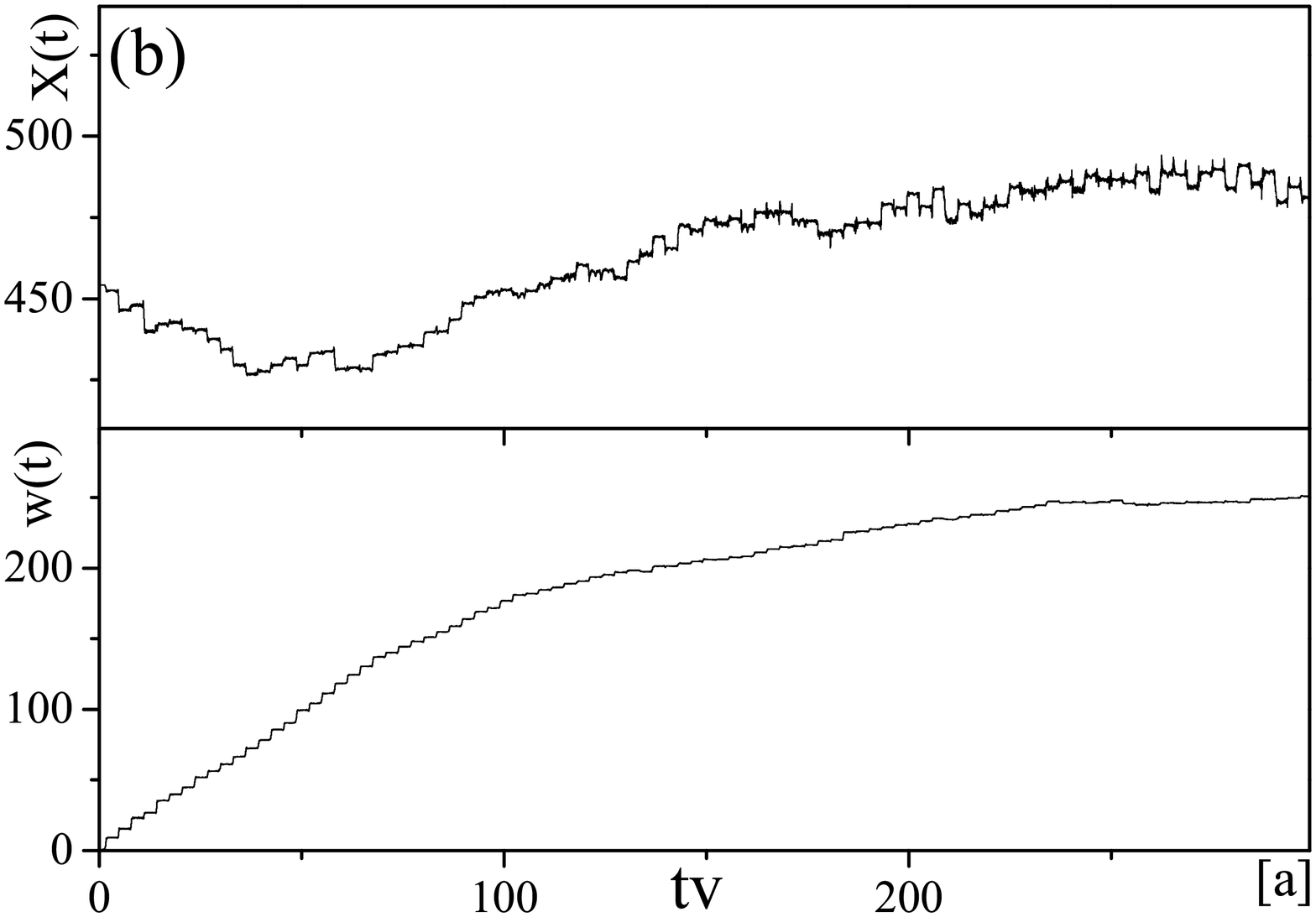}
\caption{(Color online)(a) The dynamics of the COM ($X(t)$ in the upper pannel) and width ($w(t)$ in the lower pannel) of the wavepacket in the presence of disordered potentials with (a)a periodical kicked driving and (b) a periodically modulating amplitude with the parameters  $L=1000$, $\Delta=4J$, $v=0.01Ja$.  Neither of them exhibit the sliding localized phase at low $v$.}\label{fig:SM4}
\end{figure}

\subsection{Finite size scaling of the many-body dynamics}
At half-filling case, the SLP is characterized by a persistent oscillation of the CDW order parameter $M(t)=\frac 1L \sum_i (-1)^i \langle \Psi(t)|\hat{n}_i|\Psi(t)\rangle$ if we start from the initial state $|1010\cdots10\rangle$. As for the stability of the SLP, it is important to distinguish the persistent oscillation from an extremely slow decay, and one needs to show that the oscillation amplitude doesn't decay to zero in the thermodynamic limit. To this end, we extended the simulation time of the many-body dynamics, and provided some details of the finite size scaling in this section.  We plotted the envelop of the oscillations of $M(t)$ for different system sizes up to a sufficiently long time in In Fig.\ref{fig:finitesize} (a), from which we can find that $M_{en}(t)$ barely decays in time. We define the saturated amplitude M as the average of the $M_{en}(t)$ over the period $240a/v<t<480a/v$, and study its dependence with system size. Fig.\ref{fig:finitesize} (b) shows that M decays linearly with $1/L$, and in the thermodynamic limit, M extrapolates to a finite value, which indicates that the persistent oscillation do survive in thermodynamics limit in the SLP.

\subsection{Interpolation dependence}

To recover the continuum potential function $\tilde{V}(x,t=0)$ from a set of discrete points $\{V_i(t=0)\}$, in our simulation we used the cubic spline interpolation to assure the smoothness of  the function  $\tilde{V}(x,t=0)$. Here, we will show that our results doesn't crucially depends on such a specific choice of interpolation. To this end, we choose a different interpolation method  (linear interpolation as shown in Fig.\ref{fig:SM3} a) to derive $\tilde{V}(x,t=0)$ from the same set of $\{V_i(t=0)\}$, and calculate the dynamics of the wavepacket under such a non-smooth potential. As shown in Fig.\ref{fig:SM3} (b), in spite of the roughness of the potential derived by linear interpolation, at a low moving velocity, the dynamics of the wavepacket is qualitatively the same as that in the cubic spline interpolation, thus the existence of the sliding localized phase doesn't depend on the interpolation method. However, we should emphasize that the interpolated potential should be spatially continuous, otherwise it is impossible to define a adiabatic process. For instance, for a steplike potential $\tilde{V}(x)=V_i$, if $(i-\frac 12) a<x<(i+\frac 12) a$, there is no sliding localized phase no matter how slow the potential moves. Such a steplike moving potential is equal to the periodically kicked driving as we will analyzed in the subsequential section.

\section{Other driving protocols}
\subsection{Disordered potential with a periodically kicked driving}
To the check the dependence of the SLP on the driving protocol of the disordered potential. We first consider a driving protocol with periodic kicks: instead of moving uniformly, within each driving circle, the disordered potential keeps static for a period of $\frac av$, and then is suddenly pushed forward by a lattice constant. Mathematically, this disordered potential can be expressed as
\begin{equation}
V_1(x,t)=V(x-\theta (t),0)
\end{equation}
with $\theta(t)=na$ for $(n-1)a/v<t<na/v$. Notice that on average, the disordered potential moving forward at a velocity $v$, however, the  wavepacket dynamics under such a step-like driving protocol is qualitatively different from that under an uniformly moving disordered potential. As show in Fig.\ref{fig:SM4} (a), at a low averaged velocity ($v=0.01Ja$), the COM of the wavepacket cannot follow the potential (it moves towards the opposite direction), and its width keep grows in time, indicating the absence of localization. The reason behind this is that there is no adiabatic limit for such a periodically kicked driving driving protocol no matter how small v is due to the sudden movement of the potential, thus it is impossible for the wavepacket adiabatically following the moving potential.

\subsection{Disordered potential with a periodically modulated amplitude}
 The second case considered involved a model of disordered potential with periodically modulated amplitude, in which $\tilde{V}(x,t)$ takes the form of
 \begin{equation}
 V_2(x,t)=\cos vt\times\tilde{V}(x,0)
 \end{equation}
 with $\tilde{V}(x,0)$ being defined the same as in the maintext. However, different from the moving disordered potential in the maintext, $V_2(x,t)$  preserves the discrete temporal translational symmetry:   $V_2(x,t)=V_2(x,t+2\pi/v)$. Fig.\ref{fig:SM4} (b) shows that in the presence of such a periodically driven disordered potential, there is no localizations: the width of the wavepacket keep growing even in the presence of a low frequency $v=0.01aJ$.

\begin{figure}[htb]
\includegraphics[width=0.99\linewidth,bb=70 58 961 553]{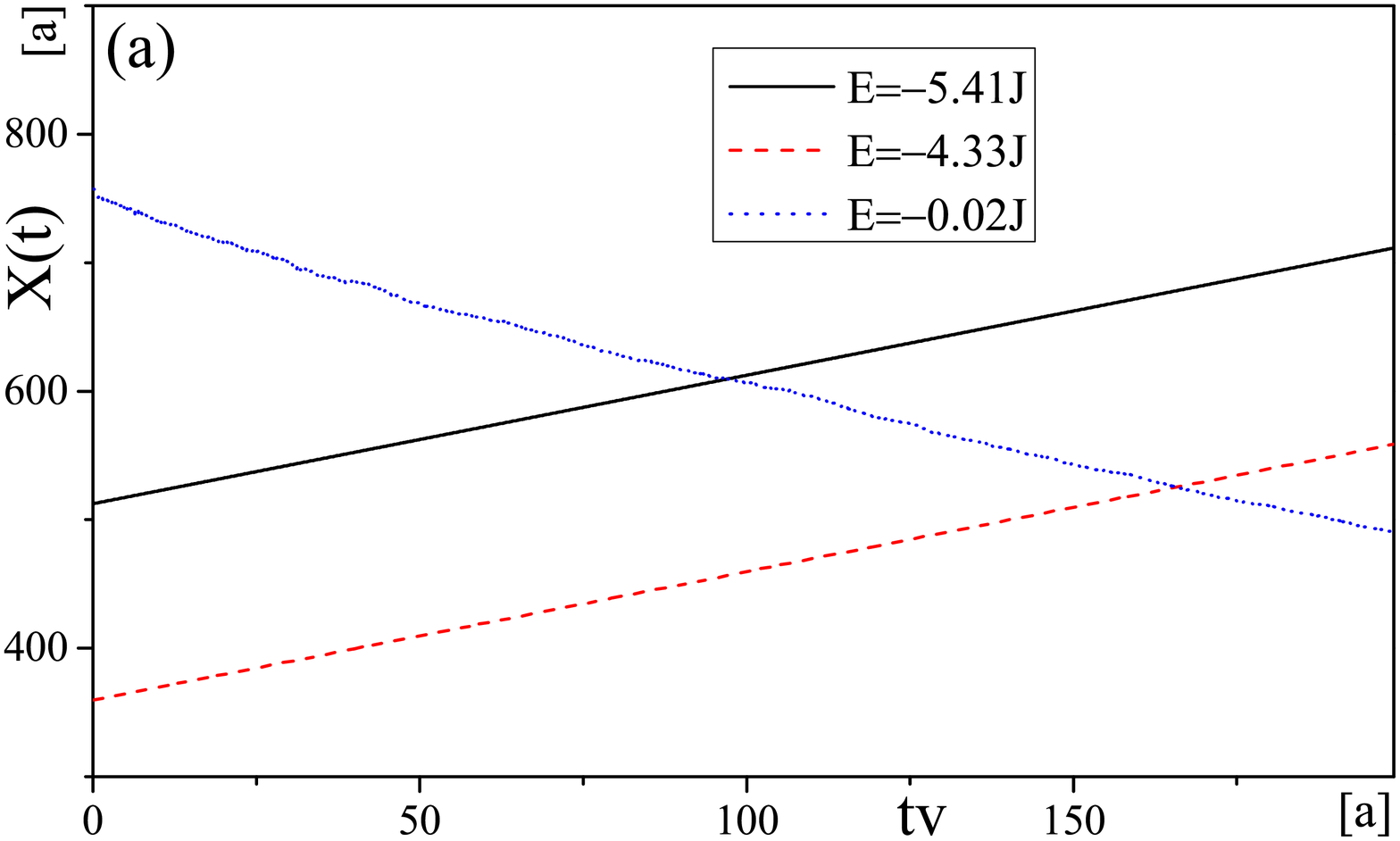}
\includegraphics[width=0.92\linewidth,bb=47 39 780 553]{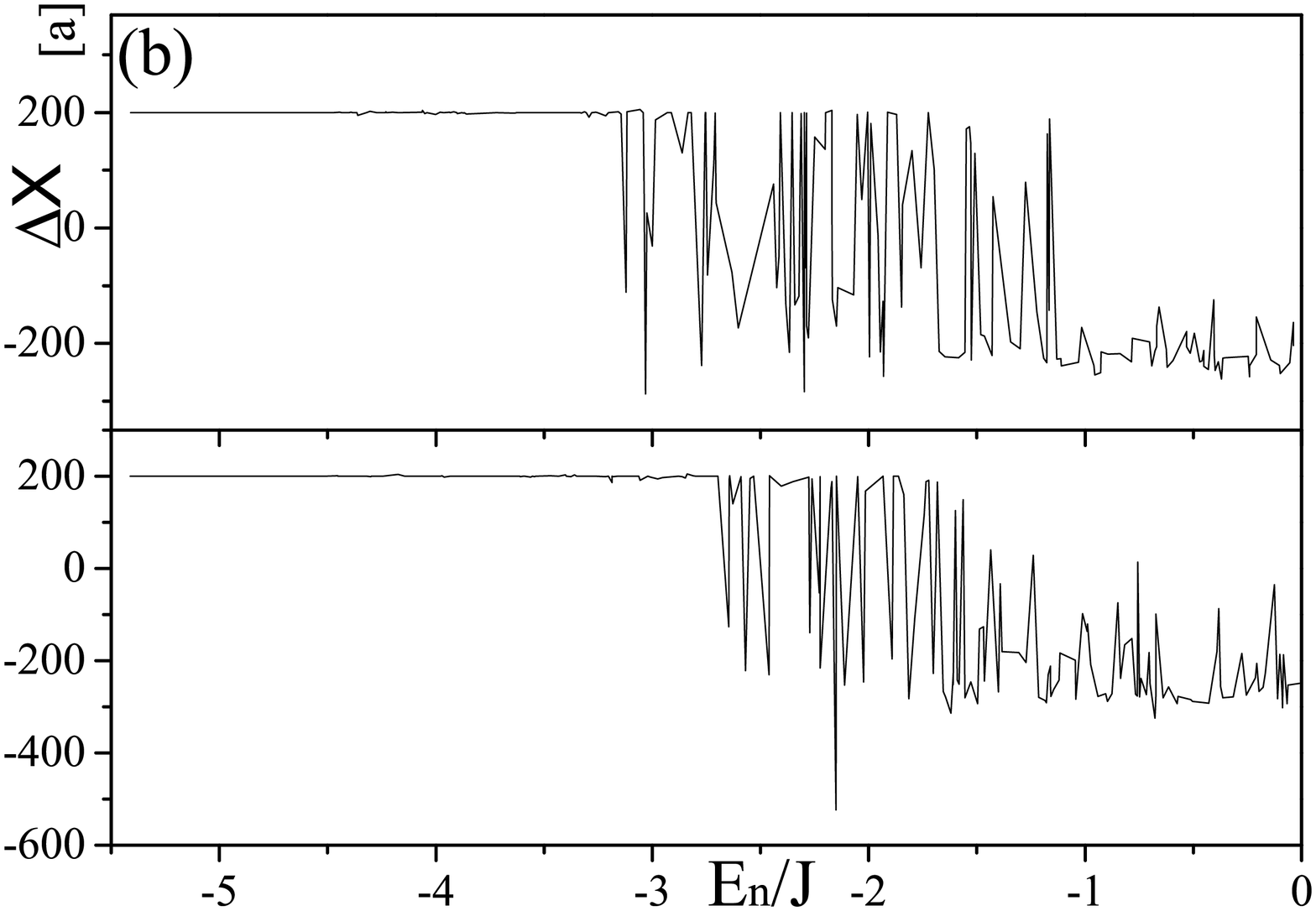}
\caption{(Color online). (a) The dynamics of the COM of the wavepacket starting from initial states with different eigen-energies for the initial state. (b) $\Delta X=X(T)-X(0)$ as a function of the initial state eigenenergy $E_n$  with parameters $Tv=200a$, $v=0.01Ja$ and $L=2000$ and two different disorder realizations (upper and lower pannels).} \label{fig:SM5}
\end{figure}

\section{Initial state dependence: the existence of a critical initial energy}

In this section, we will study the initial state dependence of the single-particle dynamics, and show that the initial state energy plays an important role in determining the long-time dynamics of the wave packet. We choose different initial states as the single-particle eigenstates of the Hamiltonian at $t=0$, all of which are spatially localized in our 1D system.  We use the energy $E_n$ to  characterize different initial states, where $E_n$ is the eigenenergy of the $n$th eigenstate. As shown in Fig.\ref{fig:SM5} (a), the long-time dynamics of the wavepacket crucially depends on $E_i$: at a fixed (low) velocity ($v=0.01aJ$), for those initial states close to the band edges (e.g. $E=-4.33J$), the long-time dynamics ($X(t)=X_0+vt$) are qualitatively identical to those starting from the ground state  ($E=-5.41J$) and we still find a sliding localized phase but with a lower critical velocity, while for those initial states close the band center (e.g. $E=-0.02J$), the COM of the wavepacket does not follow the potential (it move towards the opposite direction of the potential as shown in Fig.\ref{fig:SM5} a), and even an infinitesimal driving velocity can delocalize the system and induce a ballistic transport behavior.

For a given velocity, there  exists a critical initial energy $E_c$ to distinguish those sliding localized states  from  delocalized states. Numerically, the critical energy $E_c$ is determined by the onset of the deviation of the adiabatic COM dynamics ($X(t)=X_0+vt$), which is always accompanied by the divergence of the width of the wave packet. For a fixed $v=0.01Ja$, we plot $\Delta X=X(T)-X_0$ with $vT=200a$ as a function of the eigenenergy $E_n$ in Fig.\ref{fig:SM5} (b), from which we can find $\Delta X=200a$ for $E_n<E_c$, while for $E_n<E_c$,  $\Delta X<200a$. Fig.\ref{fig:SM5} (b) also suggests that the critical initial energy seems depends on the disorder realizations, but we are not sure whether such a dependence is a finite size effect or not.

\begin{figure}[htb]
\includegraphics[width=0.99\linewidth,bb=91 10 757 294]{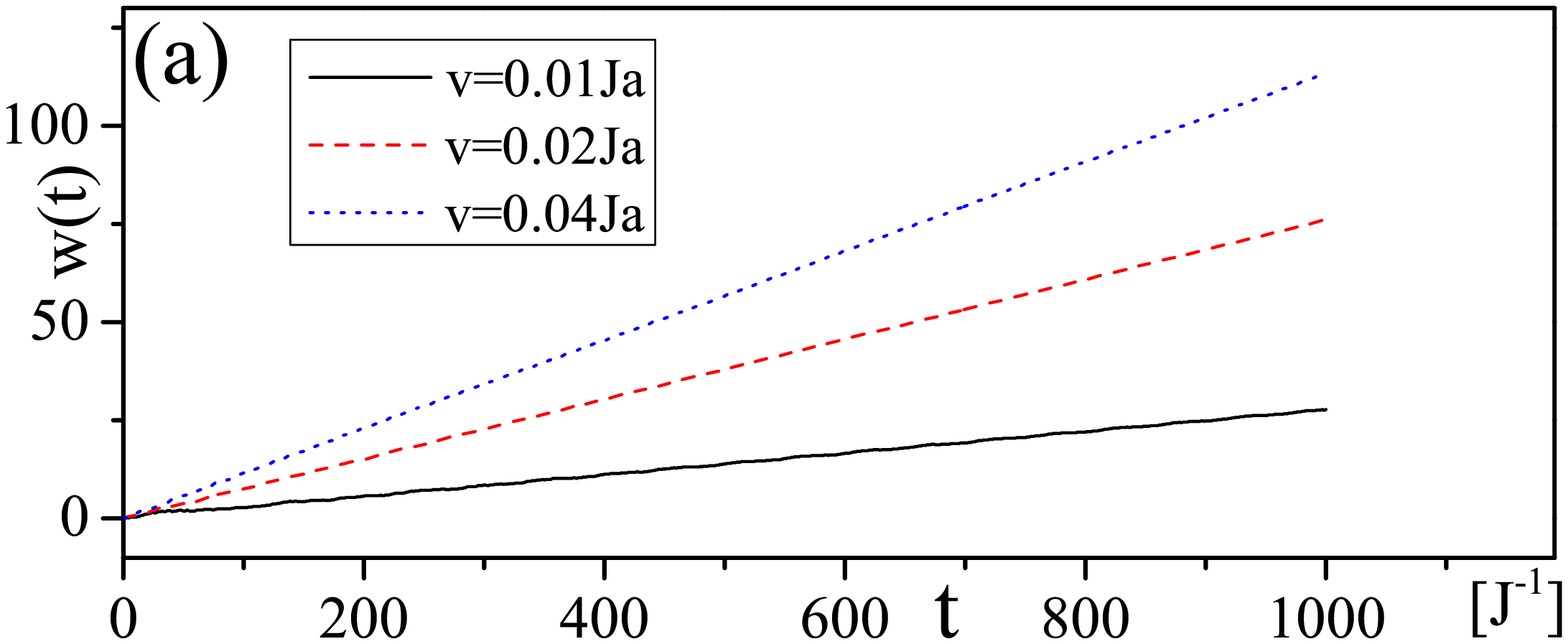}
\includegraphics[width=0.99\linewidth,bb=91 10 757 294]{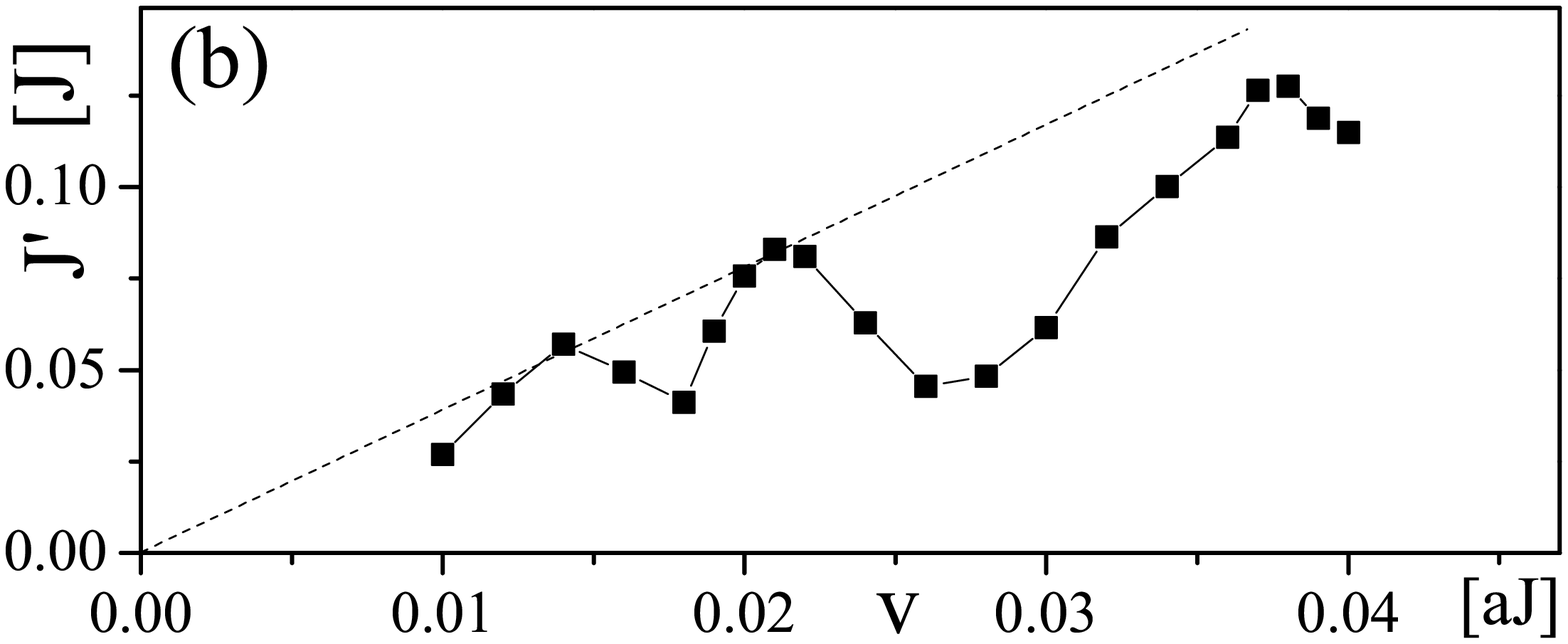}
\caption{(Color online). (a) The dynamics of width of the wavepacket in the presence of a slowly moving quasi-periodic potential $V_{k_0}(x,t)=\Delta \cos 2\pi k_0(x-vt)$ with different moving velocities; (b) the dependence of the renormalized nearest-neighboring tunneling amplitude (the slope of $w(t)$) as a function of moving velocity of the quasi-periodic potential. The parameters are chosen as $k_0=\frac{\sqrt{5}-1}2$, $\Delta=4J$, and $L=1000$.} \label{fig:SM6}
\end{figure}

\begin{figure}[htb]
\includegraphics[width=0.99\linewidth,bb=101 53 1018 574]{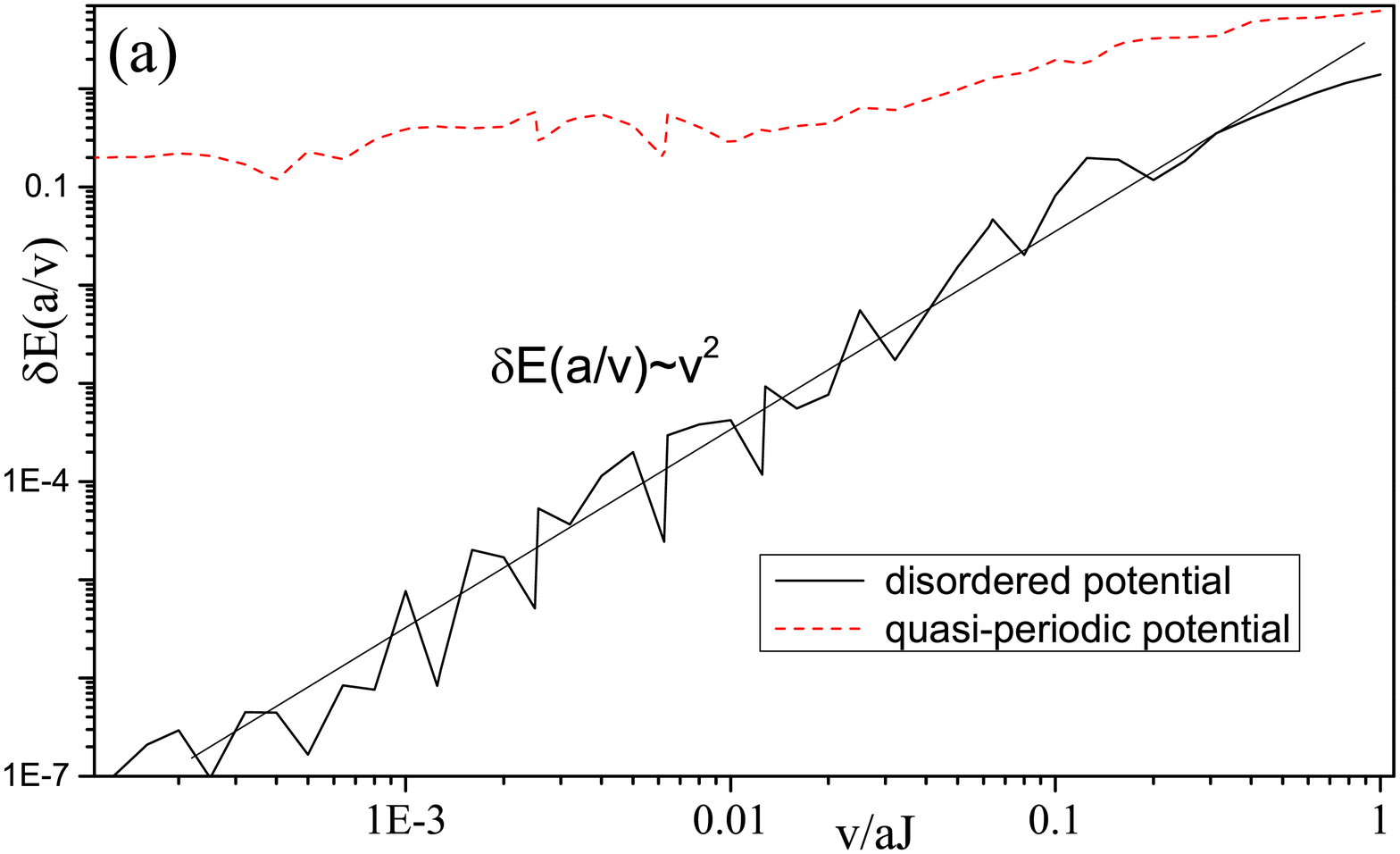}
\includegraphics[width=0.99\linewidth,bb=101 53 1018 574]{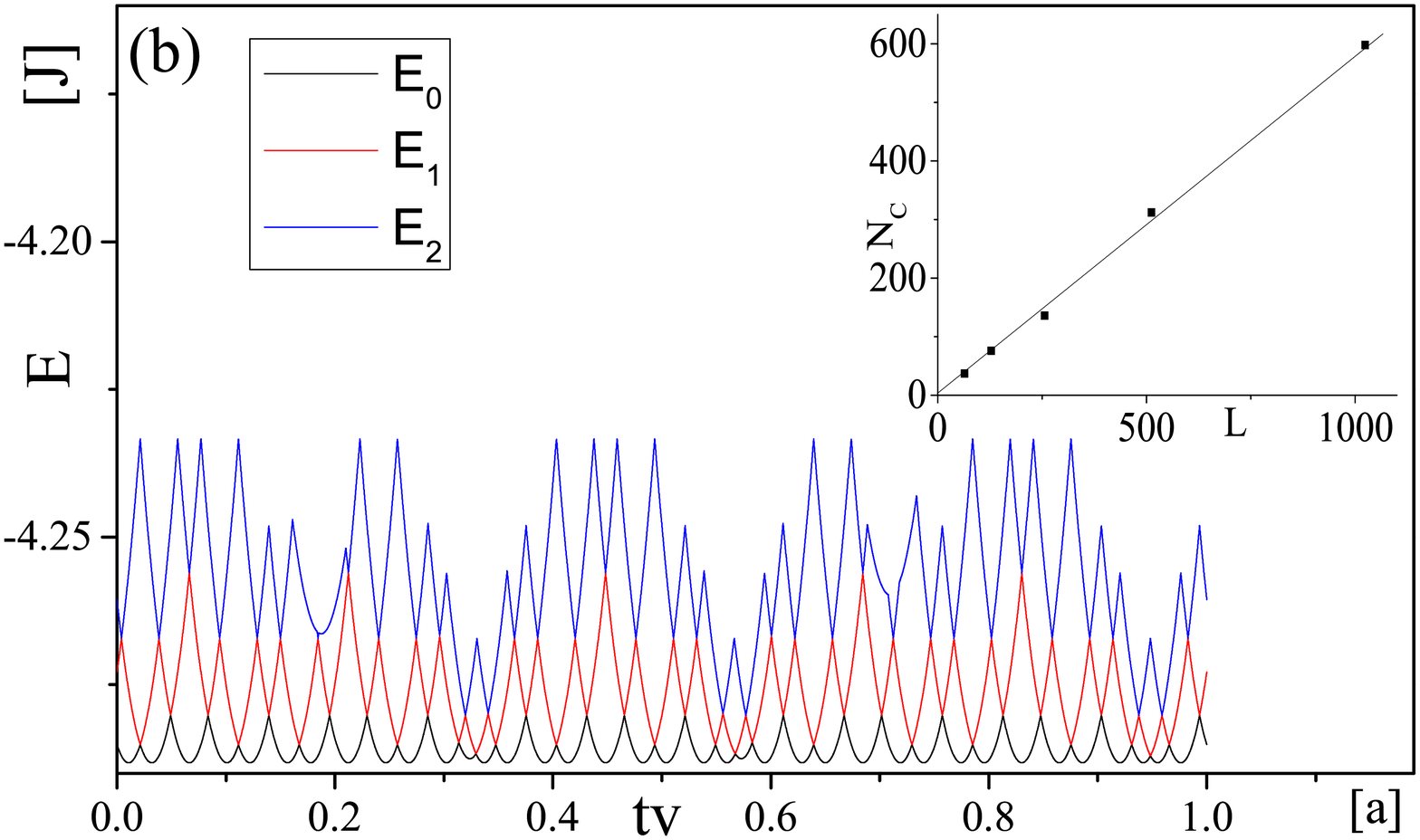}
\caption{(Color online). (a)The excess energy pumped into the system after the potential is pushed forward by one lattice constant $\delta E$ as a function of the moving velocity $v$ in the presence of disordered or quasi-periodic potentials. (b)The instantaneous eigenenergies during this period $0<tv<a$. The parameters are chosen as $L=64$ and $\Delta=4J$ for (a) and (b). The inset of (b) indicates the number of the energy crossing between the instantaneous ground state and 1st excited state as a function of the system size $L$.  } \label{fig:SM7}
\end{figure}

\section{Wavepacket dynamics in the presence of moving quasi-periodic potential}

\subsection{Ballistic transport and effective tunneling amplitude}
In the maintext, we show the wavepacket dynamics under a moving quasi-periodic potential is qualitatively different from that under a moving disordered potential, at a low velocity, the latter will be localized, while the former is characterized by a ballistic transport, where the  width of the wavepacket grows linearly with time. The slope of this linear growth can be considered as the velocity of the ballistic transport, which is proportional to the effective single-particle tunneling amplitude between adjacent sites ($J'$). As shown in Fig. \ref{fig:SM6} (a), the slopes depend on the moving velocity of the potential, from which we can derive a moving velocity dependence of the effective tunneling amplitude. Fig.\ref{fig:SM6} (b) suggests that $J'$ depends on $v$ in an oscillatory way, whose envelop roughly grows linearly with increasing $v$. In the maintext, we consider the situation where the period of the quasi-periodic potential ($a_0$) is slightly larger than the lattice constant ($a$). At a small moving velocity, the typical tunneling time between the adjacent sites is $t^*\sim \frac{a_0-a}v$, whereas the effective tunneling amplitude $J'\sim\frac 1 {t^*}$, thus is proportional to $v$. For a general $a_0$, the moving quasi-periodic potential may induce longer-range tunneling since the 1st excited state in this case may not locate at the adjacent sites of the groundstate wavepacket.

\subsection{Breakdown of adiabaticity in the presence of moving quasi-periodic potential}

In the maintext, we mentioned  that the qualitatively different behaviors between the systems with moving disordered and quasi-periodic potential is related with the breakdown of the adiabaticity in the latter, here a detailed analysis of this point is provided. In general,  for an isolated system with some external parameter being slowly driven from some initial value  to the final one with a ramp speed $v$, assuming there is no energy level crossing on the way, the adiabatic theorem states that the excess entropy or energy density ($\delta e$) pumped into the system in this process as a function of $v$  will approach to zero as $\delta e\sim v^2$  in the limit $v\rightarrow 0$. This is due to the fact that $\delta e$ is an analytic function of $v$, but insensitive to its sign ($\delta e(v)=\delta e(-v)$).

Specific to our system, the time varying parameter is the phase shift $\phi(t)$ of moving potential, which grows linearly as $\phi(t)\sim vt$, where the moving velocity $v$ is the ramp speed. For simplicity, we consider the single-particle case. Initially, $\phi(t=0)=0$,  and we choose the initial state as the ground state of $\hat{H}(t=0)$.   The final phase shift is chosen to satisfy the potential is pushed forward by one lattice constant $\phi(t=t_f)=a$, thus $t_f=a/v$. Due to the sliding space-time translational symmetry, the Hamiltonian at the final stage $\hat{H}(t=t_f)$ is equivalent to initial one $\hat{H}(t=0)$ by shifting one lattice constant ($i\rightarrow i+1$).  Under the periodical boundary condition, this equivalence indicates that the energy spectrums of $\hat{H}(t=t_f)$ and $\hat{H}(t=0)$ are the identical with each other. As a consequence, the excess energy pumped into the system in this process (the disordered/quasiperiodic potential is pushed forward by one lattice constant) is defined as:
\begin{equation}
\delta e= \langle\psi(t_f)|\hat{H}(t=t_f)|\psi(t_f)\rangle-E_0
\end{equation}
where $E_0$ is the ground state energy for both $\hat{H}(t=t_f)$ and $\hat{H}(t=0)$, which is also the system energy at $t=0$.   $|\psi(t_f)\rangle$ is the wavefunction at the final time $t_f$.

The excess energy $\delta e$ as a function of $v$ for both cases with disordered and quasi-periodic potential are plotted in Fig.\ref{fig:SM7} (a),  which indicates a qualitative difference between them: as for a moving disordered potential, the excess energy grows with the moving velocity  as $\delta e \sim v^2$, which suggests that the  adiabatical theorem still holds in this case.   On the contrary, for the moving quasi-periodic potential, $\delta e\rightarrow const$ in the limit of $v\rightarrow 0$, indicating a breakdown of the adiabatical theorem, which can explain that even an infinitesimal velocity can delocalize the system. We expect that such a breakdown of adiabatical theorem is due to the presence of a large amount of  level crossings ($\infty L$ as shown in the inset of Fig.\ref{fig:SM7} b) during this process, which pumps the particle to excited states via a sequence of Landau-Zener tunnelings.

\end{document}